\documentclass[prd,twocolumn,superscriptaddress,altaffilletter,showpacs,nofootinbib]{revtex4}
\usepackage[dvips]{graphicx}
\usepackage{amsmath}

\newcommand{\bwe}{\begin{widetext}}
\newcommand{\ewe}{\end{widetext}}
\newcommand{\be}{\begin{equation}}
\newcommand{\ee}{\end{equation}}
\newcommand{\bea}{\begin{eqnarray}}
\newcommand{\eea}{\end{eqnarray}}
\newcommand{\der}{\partial}
\newcommand{\vphi}{\varphi}

\begin{document}

\title{Brans--Dicke cosmology does not have the $\Lambda$CDM phase as an universal attractor}

\author{Ricardo Garc\'{\i}a-Salcedo}\email{rigarcias@ipn.mx}\affiliation{CICATA - Legaria del Instituto Polit\'ecnico Nacional, 11500, M\'exico, D.F., M\'exico.}

\author{Tame Gonz\'alez}\email{tamegc72@gmail.com}\affiliation{Departamento de Ingenier\'ia Civil, Divisi\'on de Ingenier\'ia, Universidad de Guanajuato, Gto., M\'exico.}

\author{Israel Quiros}\email{iquiros6403@gmail.com}\affiliation{Departamento de Ingenier\'ia Civil, Divisi\'on de Ingenier\'ia, Universidad de Guanajuato, Gto., M\'exico.}

\date{\today}

\begin{abstract}
In this paper we seek for relevant information on the asymptotic cosmological dynamics of the Brans--Dicke theory of gravity for several self-interaction potentials. By means of the simplest tools of the dynamical systems theory, it is shown that the general relativity de Sitter solution is an attractor of the Jordan frame (dilatonic) Brans--Dicke theory only for the exponential potential $U(\vphi)\propto\exp\vphi$, which corresponds to the quadratic potential $V(\phi)\propto\phi^2$ in terms of the original Brans--Dicke field $\phi=\exp\vphi$, or for potentials which asymptote to $\exp\vphi$. At the stable de Sitter critical point, as well as at the stiff-matter equilibrium configurations, the dilaton is necessarily massless. We find bounds on the Brans--Dicke coupling constant $\omega_\textsc{bd}$, which are consistent with well-known results.
\end{abstract}

\maketitle


\section{Introduction}\label{intro-sec}

The Brans--Dicke (BD) theory of gravity \cite{bd} represents the slightest modification of general relativity (GR), by the addition of a new (scalar) gravitational degree of freedom $\phi$, in addition to the 10 degrees which are associated with the metric tensor $g_{\mu\nu}$. Besides being the first physically viable modification of GR, this theory has been cornerstone for a better understanding of several other modifications of general relativity. In particular, the equivalence between BD theory and the $f(R)$ theories of gravity has been clearly established \cite{sotiriou-rev}. In contrast to Einstein's GR, the BD theory is not a fully geometrical theory of gravity. Actually, one of the propagators of the gravitational field: the metric tensor, defines the metric properties of the spacetime, meanwhile the scalar field $\phi$, which also propagates gravity, is a non-geometric field. The latter modifies the local strength of the gravitational interactions through the effective gravitational coupling $G_\text{eff}\propto\phi^{-1}$.

While many aspects of BD theory were well-explored in the past (see the textbooks \cite{maeda-book, faraoni-book}), other aspects were cleared up just recently. Thanks to the chameleon effect \cite{cham}, for instance, it was just recently understood that the lower experimental bounds on the BD coupling parameter $\omega_\textsc{bd}$, which were set up through experiments in the solar system, might not apply in the large cosmological scales if consider BD theory with a potential. According to the chameleon effect, the effective mass of the scalar field $m_\phi$ computed in the Einstein's frame, depends on the background energy density of the environment: In the large cosmological scales where the background energy density is very small (of the order of the critical density), the effective mass is also very small, so that the scalar field degree of freedom has impact in the cosmological dynamics. Meanwhile, in the solar system, where the averaged energy density of the environment is huge compared with the one in the cosmological scale, the effective mass is large, so that the Yukawa component of the gravitational interaction associated with the scalar field $\propto e^{-m_\phi r}/r$, is short-ranged, leading to an effective screening of the scalar field degree of freedom in the solar system.\footnote{In addition to the chameleon effect, the thin--shell effect, which amounts to a weakening of the effective coupling of the scalar field to the surrounding matter, conspires to allow for further screening of the scalar field in the solar system \cite{cham}.}

The Brans--Dicke theory has found very interesting applications specially in cosmology \cite{faraoni-book}, where it has been explored as a possible explanation of the present stage of the accelerated expansion of the universe \cite{bd-de}. The problem with this is that, but for some anomalies in the power spectrum of the cosmic microwave background \cite{planck-2013}, at the present stage of the cosmic evolution, any cosmological model has to approach to the so called concordance or $\Lambda$CDM model \cite{lcdm}. The mathematical basis for the latter is the GR (Einstein--Hilbert) action plus a matter action piece: 

\bea &&S_{\Lambda\text{CDM}}=\frac{1}{16\pi G}\int d^4x\sqrt{|g|}\left(R-2\Lambda\right)\nonumber\\
&&\;\;\;\;\;\;\;\;\;\;\;\;\;\;\;\;\;\;\;\;\;\;\;\;\;\;\;\;\;\;\;\;\;\;\;\;\;\;\;\;+\int d^4x\sqrt{|g|}{\cal L}_\text{CDM},\label{lcdm}\eea where ${\cal L}_\text{CDM}$ is the Lagrangian density of (pressureless) cold dark matter (CDM). On the other hand, it has been known for decades, that GR can be recovered from the BD theory only in the limit when the BD coupling constant $\omega_\textsc{bd}\rightarrow\infty$.  

In the references \cite{hrycyna, hrycyna-1, hrycyna-2}, by means of the tools of the dynamical systems theory, it was (apparently) shown that the Jordan frame (JF) Brans--Dicke theory leads naturally to the $\Lambda$CDM model since, as the authors showed in \cite{hrycyna}, the GR--de Sitter solution is an attractor of JF--BD theory, independent on the choice of the self-interaction potential for the BD scalar field.\footnote{For prior works where the de Sitter solutions are investigated within the frame of the scalar-tensor theories, see Ref. \cite{barrow-ref}, where de Sitter exact and intermediate inflationary solutions are found for FRW models with appropriate choice of the coupling function $\omega_\textsc{bd}(\vphi)$. In \cite{barrow-ref'} it is shown that intermediate ``almost de Sitter'' solutions might arise also when CDM is included. Other, more resent works on this subject, are also found \cite{odintsov-ref}.} The interesting thing is that the bounds on $\omega_\textsc{bd}$ found in \cite{hrycyna} ($\omega_\textsc{bd}\approx-3/2$), and in \cite{hrycyna-1} ($\omega_\textsc{bd}\approx-1$), are far from the solar system--based experimental bound $\omega_\textsc{bd}>40000$ \cite{will}. 

Although the chameleon effect could (in principle) explain such a discrepancy between the bounds on $\omega_\textsc{bd}$ based in solar system experimentation, and those based in cosmological considerations, nevertheless, the bounds estimated on the basis of cosmological arguments: $\omega_\textsc{bd}>120$ in \cite{aquaviva}, and $10<\omega_\textsc{bd}<10^7$ in \cite{chiva}, neither are consistent with the ones found in the references \cite{hrycyna} and \cite{hrycyna-1}. Besides, we stress that the conclusion on the existence of the GR--de Sitter (stable) critical point independent on the assumed potential in \cite{hrycyna, hrycyna-1, hrycyna-2}, is misleading. As a matter of fact, in \cite{hrycyna-2} the authors seem to recognize that the de Sitter equilibrium configuration arises only for the quadratic monomial $V(\phi)\propto\phi^2$, and for the lineal $V(\phi)\propto\phi$ potentials. Since the estimates of \cite{hrycyna, hrycyna-1, hrycyna-2} are based on the analysis of linearized solutions which are, in fact, unstable perturbations around the stable GR--de Sitter critical point and, hence, are highly dependent on the choice of the initial conditions, we suspect that these estimates could be physically meaningless.

In this paper we shall apply the simplest tools of the theory of the dynamical systems -- those which are based on knowledge of linear algebra and of the theory of the ordinary differential equations -- to uncover the dynamics of cosmological models which are based in the Brans--Dicke theory of gravity, for several self-interaction potentials, in a convenient phase space. Unlike \cite{hrycyna, hrycyna-1, hrycyna-2}, here we shall explore specific potentials other than (but also including) the quadratic and the lineal monomials: $V(\phi)\propto\phi^2$ and $V(\phi)\propto\phi$, respectively (see section \ref{pot-sec}). For a better understanding of our analysis we shall study the vacuum BD theory, and the BD theory with matter, separately (see sections \ref{vacuum-sec} and \ref{matter-sec}, respectively). In the section \ref{problem-sec}, we shall show that it is not enough that the de Sitter solution be a critical point of the dynamical system, in order for the $\Lambda$CDM model to be an attractor of the BD theory with a potential. It is a necessary condition for the latter, that the GR--de Sitter, i. e., the de Sitter point which leads to $\phi=\phi_0=const.$, to be a stable critical point instead. It will be shown that, as a matter of fact, only (exclusively) for the quadratic monomial potential $V(\phi)\propto\phi^2$, or for potentials that asymptote to $\phi^2$, the $\Lambda$CDM model is an attractor of the BD cosmology. 

Here we shall pay special attention to the discussion on the actual physical meaning -- if any -- of the linearized solutions which were used in \cite{hrycyna, hrycyna-1, hrycyna-2}, in order to bound the free parameters of the BD theory (see section \ref{discu-sec}). Our discussion favors the estimates based on computations made at the stable equilibrium configuration, over the estimates based on computations performed at the linearized solutions. For simplicity of mathematical handling we shall use the dilatonic field variable $\vphi$ instead of the standard BD field $\phi$. These variables are related by Eq. (\ref{vphi}) below.


\section{Basic setup}\label{setup-sec}

Here we assume the Brans--Dicke theory \cite{bd} to dictate the dynamics of gravity and matter. In the Jordan frame, which is the one considered in this paper, it is depicted by the following action:

\bea S=\int d^4x\sqrt{|g|}\left\{\phi R-\frac{\omega_\textsc{bd}}{\phi}(\der\phi)^2-2V+2{\cal L}_m\right\},\label{bd-action}\eea where $(\der\phi)^2\equiv g^{\mu\nu}\der_\mu\phi\der_\nu\phi$, $V=V(\phi)$ is the scalar field self-interaction potential, $\omega_\textsc{bd}$ is the BD coupling parameter, and ${\cal L}_m$ is the Lagrangian density of the matter degrees of freedom. The natural units $8\pi G=8\pi/M_\textsc{PL}^2=c=1$, are adopted. For convenience we rescale the BD scalar field and, consequently, the self-interaction potential is also redefined:

\bea \phi=e^\vphi,\;V(\phi)=e^\vphi\,U(\vphi),\label{vphi}\eea so that, the action (\ref{bd-action}) is transformed into the dilatonic BD action:

\bea S=\int d^4x\sqrt{|g|}e^\vphi\left\{R-\omega_\textsc{bd}(\der\vphi)^2-2U+2e^{-\vphi}{\cal L}_m\right\}.\label{dbd-action}\eea 

Within the context of the (low--energy) effective string theory, the latter action is meant, also, to represent the so called ``string frame'' representation of the theory \cite{wands-rev}. Here we prefer, for the moment, to keep talking about dilatonic JF--BD theory instead of string--frame effective action.

The following motion equations are obtained from (\ref{dbd-action}):

\bea &&G_{\mu\nu}=\left(\omega_\textsc{bd}+1\right)\left[\der_\mu\vphi\der_\nu\vphi-\frac{1}{2}g_{\mu\nu}(\der\vphi)^2\right]\nonumber\\
&&\;\;\;\;\;\;\;\;\;\;\;\;\;\;\;\;\;\;\;\;\;\;\;\;\;\;\;-g_{\mu\nu}\left[\frac{1}{2}(\der\vphi)^2+U(\vphi)\right]\nonumber\\
&&\;\;\;\;\;\;\;\;\;\;\;\;\;\;\;\;\;\;\;\;+\nabla_\mu\der_\nu\vphi-g_{\mu\nu}\nabla^2\vphi+e^{-\vphi}T^{(m)}_{\mu\nu},\nonumber\\
&&\nabla^2\vphi+(\der\vphi)^2=\frac{2}{3+2\omega_\textsc{bd}}\left(\der_\vphi U-U\right)\nonumber\\
&&\;\;\;\;\;\;\;\;\;\;\;\;\;\;\;\;\;\;\;\;\;\;\;\;\;\;\;\;\;\;\;\;\;\;\;\;\;\;\;\;\;\;+\frac{e^{-\vphi}}{3+2\omega_\textsc{bd}}\,T^{(m)},\label{feq}\eea where $\nabla^2\equiv g^{\mu\nu}\nabla_\mu\der_\nu$, $G_{\mu\nu}=R_{\mu\nu}-g_{\mu\nu}R/2$, and $T^{(m)}_{\mu\nu}$ is the stress-energy tensor of the matter degrees of freedom.

In this paper we shall consider Friedmann-Robertson-Walker (FRW) spacetimes with flat spatial sections for which the line-element takes the simple form: $$ds^2=-dt^2+a^2(t)\delta_{ij}dx^idx^j,\;i,j=1,2,3\,.$$ We assume the matter content of the Universe in the form of a cosmological perfect fluid, which is characterized by the following state equation $p_m=w_m\rho_m$, relating the barotropic pressure $p_m$ and the energy density $\rho_m$ of the fluid, where $w_m$ is the so called equation of state (EOS) parameter. Under these assumptions the cosmological equations (\ref{feq}) are written as it follows:

\bea &&3H^2=\frac{\omega_\textsc{bd}}{2}\,\dot\vphi^2-3H\dot\vphi+U+e^{-\vphi}\rho_m,\nonumber\\
&&\dot H=-\frac{\omega_\textsc{bd}}{2}\,\dot\vphi^2+2H\dot\vphi+\frac{\der_\vphi U-U}{3+2\omega_\textsc{bd}}\nonumber\\
&&\;\;\;\;\;\;\;\;\;\;\;\;\;\;\;-\frac{2+\omega_\textsc{bd}\left(1+w_m\right)}{3+2\omega_\textsc{bd}}\,e^{-\vphi}\rho_m,\nonumber\\
&&\ddot\vphi+3H\dot\vphi+\dot\vphi^2=2\frac{U-\der_\vphi U}{3+2\omega_\textsc{bd}}+\frac{1-3w_m}{3+2\omega_\textsc{bd}}\,e^{-\vphi}\rho_m,\nonumber\\
&&\dot\rho_m+3H\left(w_m+1\right)\rho_m=0,\label{efe}\eea where $H\equiv\dot a/a$ is the Hubble parameter.  

Due to the complexity of the system of non-linear second-order differential equations (\ref{efe}), it is a very difficult (and perhaps unsuccessful) task to find exact solutions. Yet, even when an analytic solution can be found it will not be unique but just one in a large set of them. This is in addition to the problem of the stability of given solutions. In this case the dynamical systems tools come to our rescue. These very simple tools give us the possibility to correlate such important concepts in the phase space like past and future attractors (also saddle equilibrium points), limit cycles, heteroclinic orbits, etc., with generic behavior of the dynamical system derived from the set of equations (\ref{efe}), without the need to analytically solve them. A very compact and basic introduction to the application of the dynamical systems in cosmological settings with scalar fields can be found in the references \cite{wands, coley, copeland-rev, luis, bohmer-rev, epj-2015}.


\section{dynamical systems}\label{dyn-syst-sec}

As it is for any other physical system, the possible states of a cosmological model may be also correlated with points in an equivalent state space or phase space. However, unlike in the classical mechanics case, where the phase space is spanned by the generalized coordinates and their conjugate momenta, in a cosmological context the choice of the phase space variables is not a trivial issue. This leads to a certain degree of uncertainty in the choice of an appropriate set of variables of the phase space. There are, however, certain -- not written -- rules one follows when choosing these variables: (i) these should be dimensionless variables, and (ii) whenever possible, these should be bounded. The latter requirement is necessary to have a bounded phase space where all of the existing equilibrium points are ``visible'', i. e., none of then goes to infinity. Unfortunately it is not always possible to find such bounded variables. 

Besides, a certain controversy is related with the actual usefulness of the dynamical systems approach in cosmological settings due to the apparent spurious character of attractors \cite{carroll, corichi}. In spite of this, the dynamical systems theory provides powerful tools which are commonly used in cosmology to extract essential information on the dynamical properties of a variety of cosmological models, in particular, those models where the scalar field plays a role \cite{wands, coley, copeland-rev, copeland, quiros-prd-2009, luis-mayra, luis, bohmer-rev, epj-2015, genly, cosmology-books, amendola, uggla-ref}. In the case of scalar-tensor theories, in particular the BD theory of gravity \cite{bd}, in spite of several published works \cite{ds-bd, hrycyna, hrycyna-1, hrycyna-2, holden, olga, faraoni, iranies, indios, genly-1}, the dynamical systems are not of common usage. In this paper, following the studies in \cite{hrycyna, hrycyna-1, hrycyna-2}, we want to show the power of the simplest tools of the dynamical systems theory in order to extract essential information on the cosmological dynamics of BD--based cosmological models. 

In general, when one deals with BD cosmological models it is customary to choose the following variables \cite{hrycyna, hrycyna-1, hrycyna-2}:

\bea x\equiv\frac{\dot\vphi}{\sqrt{6}H}=\frac{\vphi'}{\sqrt 6},\;y\equiv\frac{\sqrt{U}}{\sqrt{3}H},\;\xi\equiv 1-\frac{\der_\vphi U}{U},\label{vars}\eea where the tilde means derivative with respect to the variable $\tau\equiv\ln a$ -- the number of e-foldings. As a matter of fact $x$ and $y$ in Eq. (\ref{vars}), are the same variables which are usually considered in similar dynamical systems studies of FRW cosmology, within the frame of Einstein's general relativity with a scalar field matter source \cite{wands}. In terms of the above variables the Friedmann constraint in Eq. (\ref{efe}) can be written as

\bea \Omega^\text{eff}_m\equiv\frac{e^{-\vphi}\rho_m}{3H^2}=1+\sqrt{6}x-\omega_\textsc{bd}\,x^2-y^2\geq 0.\label{friedmann-c}\eea Notice that one might define a dimensionless potential energy density and an ``effective kinetic'' energy density

\bea \Omega_U=\frac{U}{3H^2}=y^2,\;\Omega^\text{eff}_K=x\left(\omega_\textsc{bd}x-\sqrt{6}\right),\label{omega-u-k}\eea respectively, so that the Friedmann constraint can be re-written in the following compact form: $$\Omega^\text{eff}_K+\Omega_U+\Omega^\text{eff}_m=1.$$ 

The definition for the dimensionless effective kinetic energy density $\Omega^\text{eff}_K$ has not the same meaning as in GR with a scalar field. It may be a negative quantity without challenging the known laws of physics. Besides, since there is not restriction on the sign of $\Omega^\text{eff}_K$, then, it might happen that $\Omega_U=U/3H^2>1$. This is due to the fact that the dilaton field in the BD theory is not a standard matter field but it is a part of the gravitational field itself. This effective (dimensionless) kinetic energy density vanishes whenever: $$x=\frac{\sqrt{6}}{\omega_\textsc{bd}}\;\Rightarrow\;\dot\vphi=\frac{6}{\omega_\textsc{bd}}\,H\;\Rightarrow\;\vphi=\frac{6}{\omega_\textsc{bd}}\,\ln a,$$ or if: $$x=0\;\Rightarrow\;\dot\vphi=0\;\Rightarrow\;\vphi=const.,$$ which, provided that the matter fluid is cold dark matter, corresponds to the GR--de Sitter universe, i. e., to the $\Lambda$CDM model. 

The following are useful equations which relate $\dot H/H^2$ and $\ddot\vphi/H^2$ with the phase space variables $x$, $y$ and $\xi$:

\bea &&\frac{\dot H}{H^2}=2\sqrt{6}\,x-3\omega_\textsc{bd}\,x^2-\frac{3y^2\xi}{3+2\omega_\textsc{bd}}\nonumber\\
&&\;\;\;\;\;\;\;\;\;\;\;\;\;\;\;\;\;\;\;\;\;\;\;\;\;\;\;\;\;\;-\frac{2+\omega_\textsc{bd}\left(1+w_m\right)}{3+2\omega_\textsc{bd}}\,3\Omega^\text{eff}_m,\nonumber\\
&&\frac{\ddot\vphi}{H^2}=-3\sqrt{6}\,x-6x^2+\frac{6y^2\xi}{3+2\omega_\textsc{bd}}\nonumber\\
&&\;\;\;\;\;\;\;\;\;\;\;\;\;\;\;\;\;\;\;\;\;\;\;\;\;\;\;\;\;\;\;\;\;\;\;\;\;\;\;\;\;\;\;\;+\frac{1-3w_m}{3+2\omega_\textsc{bd}}\,3\Omega^\text{eff}_m.\label{useful}\eea

Our goal will be to write the resulting system of cosmological equations (\ref{efe}), in the form of a system of autonomous ordinary differential equations (ODE-s) in terms of the variables $x$, $y$, $\xi$, of some phase space. We have:

\bea &&x'=\frac{\ddot\vphi}{\sqrt{6}H^2}-x\frac{\dot H}{H^2},\nonumber\\
&&y'=y\left[\frac{\sqrt 6}{2}\left(1-\xi\right)x-\frac{\dot H}{H^2}\right],\nonumber\\
&&\xi'=-\sqrt{6}x\left(1-\xi\right)^2\left(\Gamma-1\right),\;\Gamma\equiv\frac{U\der^2_\vphi U}{(\der_\vphi U)^2},\label{ode'}\eea or, after substituting equations (\ref{useful}) into (\ref{ode'}), we obtain the following autonomous system of ODE-s:

\bea &&x'=-3x\left(1+\sqrt{6}x-\omega_\textsc{bd}x^2\right)+\frac{x+\sqrt{2/3}}{3+2\omega_\textsc{bd}}\,3y^2\xi\nonumber\\
&&\;\;\;\;\;\;\;\;\;\;\;\;\;\;+\frac{\frac{1-3w_m}{\sqrt 6}+\left[2+\omega_\textsc{bd}(1+w_m)\right]\,x}{3+2\omega_\textsc{bd}}\,3\Omega^\text{eff}_m,\nonumber\\
&&y'=y\left[3x\left(\omega_\textsc{bd}x-\frac{\xi+3}{\sqrt{6}}\right)+\frac{3y^2\xi}{3+2\omega_\textsc{bd}}\right.\nonumber\\
&&\left.\;\;\;\;\;\;\;\;\;\;\;\;\;\;\;\;\;\;\;\;\;\;\;\;\;\;\;\;+\frac{2+\omega_\textsc{bd}\left(1+w_m\right)}{3+2\omega_\textsc{bd}}\,3\Omega^\text{eff}_m\right],\nonumber\\
&&\xi'=-\sqrt{6}x\left(1-\xi\right)^2\left(\Gamma-1\right),\label{asode}\eea where $\Omega^\text{eff}_m$ is given by Eq. (\ref{friedmann-c}), and it is assumed that $\Gamma=U\der^2_\vphi U/(\der_\vphi U)^2$ can be written as a function of $\xi$ \cite{epj-2015}: $\Gamma=\Gamma(\xi)$. Hence, the properties of the dynamical system (\ref{asode}) are highly dependent on the specific functional form of the potential $U=U(\vphi)$.


\section{the dynamical system for different self-interaction potentials}\label{pot-sec}

In this section we shall write the dynamical system (\ref{asode}) for a variety of self-interaction potentials of cosmological interest. It is worth noticing that the only information on the functional form of the self-interaction potential is encoded in the definition of the parameter $\Gamma$ in Eq. (\ref{asode}). Hence, what we need is to write the latter parameter as a concrete function of the coordinate $\xi$ for given potentials.

\subsection{Exponential potential}\label{sub-exp}

We start with the study of the exponential potential

\bea U(\vphi)=M^2\,e^{k\vphi},\label{exp}\eea which, in terms of the standard BD field $\phi$ (see Eq. (\ref{vphi})), amounts to the power-law potential $V(\phi)=M^2\phi^{k+1}$ in the action (\ref{bd-action}). In Eq. (\ref{exp}), $M^2$ and $k$ are free constant parameters. In this -- the most simple -- case $$\xi=1-\frac{\der_\vphi U}{U}=1-k,$$ is a constant, so that the system of ODE-s (\ref{asode}) reduces dimensionality from 3 to 2. The fact that, for the exponential potential $\Gamma=1$, is unimportant in this case since, as said, $\xi$ is not a variable but a constant. 

The following plane-autonomous system of ODE-s is obtained:

\bea &&x'=-3x\left(1+\sqrt{6}x-\omega_\textsc{bd}x^2\right)\nonumber\\
&&\;\;\;\;\;\;\;\;\;\;\;\;\;\;\;\;\;\;\;\;\;\;\;\;\;\;\;\;+\frac{3(1-k)}{3+2\omega_\textsc{bd}}\left(x+\sqrt{2/3}\right)y^2\nonumber\\
&&\;\;\;\;\;\;\;\;\;\;\;+\frac{\frac{1-3w_m}{\sqrt 6}+\left[2+\omega_\textsc{bd}(1+w_m)\right]\,x}{3+2\omega_\textsc{bd}}\,3\Omega^\text{eff}_m,\label{x-ode-1}\\
&&y'=y\left[3x\left(\omega_\textsc{bd}x-\frac{4-k}{\sqrt 6}\right)+\frac{3(1-k)}{3+2\omega_\textsc{bd}}\,y^2\right.\nonumber\\
&&\left.\;\;\;\;\;\;\;\;\;\;\;\;\;\;\;\;\;\;\;\;\;\;\;\;\;\;\;\;\;+\frac{2+\omega_\textsc{bd}\left(1+w_m\right)}{3+2\omega_\textsc{bd}}\,3\Omega^\text{eff}_m\right],\label{y-ode-1}\eea where, as stated before, $\Omega^\text{eff}_m$ is given by Eq. (\ref{friedmann-c}).

\subsection{Combination of exponentials}\label{sub-comb-exp}

The combination of exponentials ($M^2$, $N^2$, $k$ and $m$ are free constant parameters):

\bea U(\vphi)=M^2\,e^{k\vphi}+N^2\,e^{m\vphi},\label{comb-exp}\eea which corresponds to the BD potential $$V(\phi)=M^2\phi^{k+1}+N^2\phi^{m+1},$$ leads to the following

\bea \Gamma(\xi)=(k+m)\frac{\left(1-\frac{mk}{k+m}-\xi\right)}{\left(1-\xi\right)^2},\label{gamma-comb-exp}\eea where we have taken into consideration that $$\xi=1-\frac{\der_\vphi U}{U}=\frac{1-k+(1-m)\left(\frac{N}{M}\right)^2 e^{(m-k)\vphi}}{1+\left(\frac{N}{M}\right)^2 e^{(m-k)\vphi}}.$$ As a consequence the third autonomous ODE in the dynamical system (\ref{asode}) can be written as

\bea \xi'=-\sqrt{6}\,x\,\left[k+m-mk-1-(k+m-2)\xi-\xi^2\right].\label{xi-ode-comb-exp}\eea

The particular case when $M^2=N^2$, $m=-k$, corresponds to the cosh potential:

\bea U(\vphi)=2M^2\cosh(k\vphi),\label{cosh}\eea for which $\Gamma(\xi)=k^2/(1-\xi)^2$, and 

\bea \xi'=-\sqrt{6}\,x\,\left[k^2-(1-\xi)^2\right].\label{xi-ode-cosh}\eea

\subsection{cosh and sinh-like potentials}\label{sub-cosh-sinh}

The cosh-like potentials

\bea U(\vphi)=M^2\cosh^k(\mu\vphi),\label{cosh-like}\eea where $M^2$, $k$ and $\mu$ are constant parameters, are also very interesting from the point of view of the cosmology \cite{cosh-pot}. These correspond to potentials of the following kind

\bea V(\phi)=M^2 \phi\left[\cosh(\ln\phi^\mu)\right]^k,\label{cosh-like-bd}\eea in terms of the original BD field $\phi$. We have $$\xi=1-\frac{\der_\vphi U}{U}=1-k\mu\tanh(\mu\vphi),$$ so that 

\bea \Gamma(\xi)=\frac{k^2\mu^2+(k-1)(1-\xi)^2}{k(1-\xi)^2}.\label{gamma-cosh-like}\eea The resulting autonomous ODE -- third equation in (\ref{asode}) -- reads

\bea \xi'=-\frac{\sqrt{6}}{k}\,x\left[k^2\mu^2-(1-\xi)^2\right].\label{xi-ode-cosh-like}\eea Notice that, by setting $k=1$ and then replacing $k\rightarrow\mu$ one recovers the ODE (\ref{xi-ode-cosh}) for the cosh potential (\ref{cosh}).

Working in a similar way with the sinh-like potential

\bea U(\vphi)=M^2\sinh^k(\mu\vphi),\label{sinh-like}\eea we obtain: $$\xi=1-k\mu\,\text{cotanh}(\mu\vphi),$$ and the same $$\Gamma(\xi)=\frac{k^2\mu^2+(k-1)(1-\xi)^2}{k(1-\xi)^2},$$ so that the corresponding autonomous ODE is the same Eq. (\ref{xi-ode-cosh-like}) as for the cosh-like potential. The difference resides in the range of the variable $\xi$. For the cosh-like potential one has:

\bea 1-k\mu\leq\xi\leq 1+k\mu\;(-\infty<\vphi<\infty),\label{xi-range-cosh-like}\eea while, for the sinh-like one $$1+k\mu\leq\xi<\infty,$$ when $-\infty<\vphi<0$, and $$-\infty<\xi\leq 1-k\mu,$$ if $0<\vphi<\infty$. Here we have assumed that both $k$ and $\mu$ are non-negative quantities ($k\geq 0$, $\mu\geq 0$).


\section{vacuum brans-dicke cosmology}\label{vacuum-sec}

A significant simplification of the dynamical equations is achieved when matter degrees of freedom are not considered. In this case, since $\Omega^\text{eff}_m=0\;\Rightarrow\;y^2=1+\sqrt{6}x-\omega_\textsc{bd}\,x^2,$ then the system of ODE-s (\ref{asode}) simplifies to a plane-autonomous system of ODE-s:

\bea &&x'=\left(-3x+3\frac{x+\sqrt{2/3}}{3+2\omega_\textsc{bd}}\,\xi\right)\left(1+\sqrt{6}x-\omega_\textsc{bd}x^2\right),\nonumber\\
&&\xi'=-\sqrt{6}x\left(1-\xi\right)^2\left(\Gamma-1\right).\label{x-xi-ode-vac}\eea 

In the present case one has

\bea &&\Omega_U=\frac{U}{3H^2}=y^2=1+\sqrt{6}x-\omega_\textsc{bd}x^2,\nonumber\\
&&\Omega^\text{eff}_K=x\left(\omega_\textsc{bd}x-\sqrt{6}\right)\;\Rightarrow\;\Omega^\text{eff}_K+\Omega_U=1,\label{omegas-vac}\eea where we recall that the definition of the effective (dimensionless) kinetic energy density $\Omega^\text{eff}_K$, has not the same meaning as in GR with scalar field matter, and it may be, even, a negative quantity. In this paper we consider non-negative self-interaction potentials $U(\vphi)\geq 0$, so that the dimensionless potential energy density $\Omega_U=y^2$, is restricted to be always non-negative: $\Omega_U=1+\sqrt{6}x-\omega_\textsc{bd}x^2\geq 0$. Otherwise, $y^2<0$, and the phase-plane would be a complex plane. Besides, we shall be interested in expanding cosmological solutions exclusively ($H\geq 0$), so that $y\geq 0$. Because of this the variable $x$ is bounded to take values within the following interval:

\bea \alpha_-\leq x\leq\alpha_+,\;\alpha_\pm=\sqrt\frac{3}{2}\left(\frac{1\pm\sqrt{1+2\omega_\textsc{bd}/3}}{\omega_\textsc{bd}}\right).\label{x-bound}\eea This means that the phase space for the vacuum Brans--Dicke theory $\Psi_\text{vac}$ can be defined as follows:

\bea &&\Psi_\text{vac}=\left\{(x,\xi):\;\alpha_-\leq x\leq\alpha_+\right\},\label{vac-phase-space}\eea where the bounds on the variable $\xi$ --  if any -- are set by the concrete form of the self-interaction potential (see below).

Another useful quantity is the deceleration parameter 

\bea &&q=-1-\frac{\dot H}{H^2}=-1-2\sqrt{6}x+3\omega_\textsc{bd}x^2\nonumber\\
&&\;\;\;\;\;\;\;\;\;\;\;\;\;\;\;\;\;\;\;\;\;\;\;\;\;\;+\frac{3(1+\sqrt{6}x-\omega_\textsc{bd}x^2)\xi}{3+2\omega_\textsc{bd}}.\label{dec-par-zero-m}\eea

Seemingly, in accordance with the results of \cite{hrycyna, hrycyna-1, hrycyna-2}, without the specification of the function $\Gamma(\xi)$, there are found four dilatonic equilibrium points $P_i:(x_i,\xi_i)$, in the phase space corresponding to the dynamical system (\ref{x-xi-ode-vac}). The first one is the GR--de Sitter phase: 

\bea &&(0,0)\;\Rightarrow\;x=0\;\Rightarrow\;\vphi=\vphi_0,\;\text{and}\nonumber\\
&&\;\;\;\;\;\;\;\;\;\;\;\;\;\;\;y^2=1\;\Rightarrow\;3H^2=U=const.,\nonumber\eea which corresponds to accelerated expansion $q=-1$. Given that, the eigenvalues of the linearization matrix around this point depend on the concrete form of the function $\Gamma(\xi)$, $$\lambda_{1,2}=-\frac{3}{2}\left(1\pm\sqrt{1+\frac{8(1-\Gamma)}{3(3+2\omega_\textsc{bd})}}\right),$$ at first sight it appears that nothing can be said about the stability of this solution until the functional form of the self-interaction potential is specified. Notice, however, that since $\xi=0$ at this equilibrium point, this means that $U(\vphi)\propto e^\vphi$, i. e., the function $\Gamma$ is completely specified: $\Gamma=1$. As a matter of fact, the eigenvalues of the linearization matrix around $(0,0)$ are: $\lambda_1=-3,\;\lambda_2=0.$ This means that $(0,0)$ is a non-hyperbolic point.

We found, also, another de Sitter solution: $q=-1$ $\Rightarrow\;\dot H=0$, which is associated with scaling of the effective kinetic and potential energies of the dilaton:

\bea &&P:\left(\frac{1}{\sqrt{6}(1+\omega_\textsc{bd})},1\right)\;\Rightarrow\nonumber\\
&&\frac{\Omega^\text{eff}_K}{\Omega_U}=-\frac{6+5\omega_\textsc{bd}}{12+17\omega_\textsc{bd}+6\omega^2_\textsc{bd}},\nonumber\\
&&\lambda_1=-\frac{4+3\omega_\textsc{bd}}{1+\omega_\textsc{bd}},\;\lambda_2=0,\label{dil-scaling}\eea where, as before, $\lambda_1$ and $\lambda_2$ are the eigenvalues of the linearization matrix around the critical point. We call this as BD--de Sitter critical point to differentiate it from the GR--de Sitter point. 

In order to make clear what the difference is between both de Sitter solutions, let us note that the Friedmann constraint (\ref{friedmann-c}), evaluated at the BD--de Sitter point above, can be written as $$e^{-\vphi}\rho_m=3H_0^2+\frac{6+5\omega_\textsc{bd}}{6(1+\omega_\textsc{bd})^2}\,3H_0^2-U_0,$$ i. e., $e^{-\vphi}\rho_m=const.$ This means that the weakening/strengthening of the effective gravitational coupling ($G_\text{eff}\propto e^{-\vphi}$) is accompanied by a compensating growing/decreasing property of the energy density of matter $\rho_m\propto e^\vphi$, which leads to an exponential rate o expansion $a(t)\propto e^{H_0 t}$. This is to be contrasted with the GR--de Sitter solution: $3H_0^2=U_0$ $\Rightarrow\;a(t)\propto e^{\sqrt{U_0/3}\,t}$, which is obtained only for vacuum, $\rho_\text{vac}=U_0$; $\rho_m=0$.

The effective stiff-dilaton critical points ($\Omega^\text{eff}_K=1$):

\bea &&P_\pm:\left(\alpha_\pm,1\right)\;\Rightarrow\;q_\pm=2+\sqrt{6}\,\alpha_\pm,\nonumber\\
&&\lambda^\pm_1=6\left(1+\sqrt\frac{2}{3}\,\alpha_\pm\right),\;\lambda_2=0,\label{stiff-dil-vac}\eea are also found, where the $\alpha_\pm$ are defined in Eq. (\ref{x-bound}).

In the paragraph starting below equation (\ref{dec-par-zero-m}), we said that, seemingly (in accordance with the results of the references \cite{hrycyna, hrycyna-1, hrycyna-2}), the obtained critical points are quite independent of the form of the function $\Gamma$. Notice, however, that this is not true at all. For the GR--de Sitter point, for instance, $\xi=0$, which means that $$\xi=1-\frac{\der_\vphi U}{U}=0\;\Rightarrow\;U\propto e^\vphi,$$ forcing $\Gamma=1$. For the remaining equilibrium points, $\xi=1$ $\Rightarrow\;U=const$, and $\Gamma=$undefined. This means that the equilibrium points listed above exist only for specific self--interaction potentials, but not for arbitrary potentials. Hence, contrary to the related statements in \cite{hrycyna, hrycyna-1, hrycyna-2}, the above results are not as general as they seem to be.

Given that the critical points obtained before were all non-hyperbolic, resulting in a lack of information on the corresponding asymptotic properties, in the following subsections we shall focus in the exponential potential (\ref{exp}): $U(\vphi)\propto\exp(k\vphi)$ $\Rightarrow\;\xi=1-k$, which includes the particular case when $$k=1\;\Rightarrow\;\xi=0\;\Rightarrow\;U(\vphi)=M^2\exp\vphi\;\Rightarrow\;\Gamma=1,$$ and the cosmological constant case $$k=0\;\Rightarrow\;\xi=1\;\Rightarrow\;U=M^2,$$ with the hope to get more precise information on the stability properties of the corresponding equilibrium configurations.\footnote{When the critical point under scrutiny is a non-hyperbolic point the linear analysis is not enough to get useful information on the stability of the point. In this case other tools, such as the center manifold theorem \cite{classic-books} are to be invoked.} These particular cases: $\xi=0$, and $\xi=1$, correspond to the four critical points obtained above. For completeness we shall consider also other potentials than the exponential.

\subsection{Exponential potential}

Let us investigate the vacuum FRW--BD cosmology driven by the exponential potential (\ref{exp}). In this case, since $\xi=1-k$, is a constant, the plane-autonomous system of ODE-s (\ref{x-xi-ode-vac}), simplifies to a single autonomous ODE:

\bea &&x'=-\left(\frac{\left(k+2+2\omega_\textsc{bd}\right)x-\sqrt\frac{2}{3}(1-k)}{1+2\omega_\textsc{bd}/3}\right)\times\nonumber\\
&&\;\;\;\;\;\;\;\;\;\;\;\;\;\;\;\;\;\;\;\;\;\;\;\;\;\;\;\;\;\;\;\;\times\left(1+\sqrt{6}x-\omega_\textsc{bd}x^2\right).\label{x-ode-vac-exp}\eea 

The critical points of the latter dynamical system are:

\bea x_1=\frac{\sqrt{2/3}\,(1-k)}{k+2+2\omega_\textsc{bd}},\;x_\pm=\alpha_\pm,\label{vac-exp-c-points}\eea where the $\alpha_\pm$ are given by Eq. (\ref{x-bound}). Notice that, since $x_i\neq 0$ (but for $k=1$, in which case $x_1=0$ and $q=-1$), there are not critical points associated with constant $\vphi=\vphi_0$. This means that the de Sitter phase with $\dot\vphi=0$ ($\vphi=const$), $U(\vphi)=const.$, i. e., the one which occurs in GR and which stands at the heart of the $\Lambda$CDM model, does not arise in the general case when $k\neq 1$. 

Hence, only in the particular case of the exponential potential (\ref{exp}) with $k=1$ ($\xi=0$), which corresponds to the quadratic potential in terms of the original BD variables: $V(\phi)=M^2\phi^2$, the GR-de Sitter phase is a critical point of the dynamical system (\ref{x-ode-vac-exp}). In this case the critical points are (see Eq. (\ref{vac-exp-c-points})): $x_1=0$, $x_\pm=\alpha_\pm$. Worth noticing that $x_1=0$ corresponds to the GR--de Sitter solution $3H^2=M^2\exp\vphi_0$, meanwhile, the $x_\pm=\alpha_\pm$, correspond to the stiff-fluid (kinetic energy) dominated phase: $\Omega^\text{eff}_K=1$. While in the former case the deceleration parameter $q=-1-\dot H/H^2=-1$, in the latter case it is found to be

\bea q=2+\sqrt{6}\,\alpha_+>0.\label{dec-p-stiff}\eea

For small (linear) perturbations $\epsilon=\epsilon(\tau)$ around the critical points: $x=x_i+\epsilon$, $\epsilon\ll 1$, one has that, around the de Sitter solution: $\epsilon'=-3\epsilon$ $\Rightarrow\;\epsilon(\tau)\propto\exp(-3\tau)$, so that it is an attractor solution. Meanwhile, around the stiff-matter solutions: $$\epsilon_\pm(\tau)\propto e^{3\left(2+\sqrt{6}\,\alpha_\pm\right)\tau},$$ so that, if assume non-negative $\omega_\textsc{bd}\geq 0$, the points $x_\pm$ are always past attractors (unstable equilibrium points) since $2+\sqrt{6}\,\alpha_->0$. For negative $\omega_\textsc{bd}<0$, these points are both past attractors whenever $\omega_\textsc{bd}<-3/2$. In this latter case, for $-3/2<\omega_\textsc{bd}<0$, the point $x_+$ is a past attractor, while the point $x_-$ is a future attractor instead.

\subsection{Constant potential $U(\vphi)=M^2$}\label{cc}

The constant potential is a particular case of the exponential (\ref{exp}), when $k=0$ ($\xi=1$ $\Rightarrow\;U=const$). In this case the autonomous ODE (\ref{x-ode-vac-exp}) simplifies:

\bea x'=\left[\frac{\sqrt{2/3}-2(1+\omega_\textsc{bd})x}{3+2\omega_\textsc{bd}}\right]\left(1+\sqrt{6}x-\omega_\textsc{bd}x^2\right).\label{x-ode-vac-cc}\eea The critical points correspond to the following values of the independent variable $x$:

\bea x_1=\frac{1}{\sqrt{6}(1+\omega_\textsc{bd})},\;x_\pm=\alpha_\pm.\label{c-points-cc}\eea Since, in this case, 

\bea &&\frac{\dot H}{H^2}=-\frac{3-\sqrt{6}\omega_\textsc{bd}x}{3+2\omega_\textsc{bd}}\left[1-\sqrt{6}(1+\omega_\textsc{bd})x\right]\nonumber\\
&&\left.\;\;\;\;\;\;\;\;\;\;\;\;\Rightarrow\;\frac{\dot H}{H^2}\right|_{x_1}=0\;\Rightarrow\;H=H_0,\label{hdot}\eea the point $x_1$ corresponds to BD--de Sitter expansion ($q=-1$). At $x_1$ the effective kinetic and potential energies of the dilaton scale as $$\frac{\Omega^\text{eff}_K}{\Omega_U}=-\frac{6+5\omega_\textsc{bd}}{12+17\omega_\textsc{bd}+6\omega^2_\textsc{bd}},$$ where, as mentioned before, the minus sign is not problematic since $\Omega^\text{eff}_K$ is not the kinetic energy of an actual matter field. As already shown -- see the paragraph starting below Eq. (\ref{dil-scaling}) and ending above Eq. (\ref{stiff-dil-vac}) -- this point does not correspond to a $\Lambda$CDM phase of the cosmic evolution, since, unlike in the GR case, in the BD theory the effective gravitational coupling $G_\text{eff}\propto e^{-\vphi}$ is not a constant and, besides, the de Sitter solution $H=H_0$ is obtained in the presence of ordinary matter with energy density $\rho_m\propto G^{-1}_\text{eff}$. For more on this see section \ref{problem-sec}.

Given that under a small perturbation ($\epsilon\ll 1$) around $x_1$: $$\epsilon(\tau)\propto\exp\left(-\frac{4+3\omega_\textsc{bd}}{1+\omega_\textsc{bd}}\,\tau\right),$$ this is a stable equilibrium point (future attractor) if the BD parameter $\omega_\textsc{bd}\geq 0$. In case it were a negative quantity, instead, $x_1$ were a future attractor whenever $\omega_\textsc{bd}<-4/3$ and $-1<\omega_\textsc{bd}<0$.

The critical points $x_\pm$ in Eq. (\ref{c-points-cc}), correspond to kinetic energy--dominated phases, i. e., to stiff-matter solutions $\Omega^\text{eff}_K=1$, where $q=2+\sqrt{6}\,\alpha_+>0$, and, under a small perturbation $\epsilon'=\lambda_\pm\epsilon$, $$\lambda_\pm=6\left(1+\sqrt\frac{2}{3}\,\alpha_\pm\right),$$ so that, assuming non-negative $\omega_\textsc{bd}\geq 0$, the points $x_\pm$ are always unstable (source critical points). In the case when $\omega_\textsc{bd}<0$ is a negative quantity, the point $x_-$ is unstable if $\omega_\textsc{bd}<-4/3$ (the critical point $x_+$ is always unstable).

\subsection{Other potentials than the exponential}\label{other}

The concrete form of the dynamical system (\ref{x-xi-ode-vac}) depends crucially on the function $\Gamma(\xi)$. For a combination of exponentials, for instance, one has (see Eq. (\ref{gamma-comb-exp})):

\bea &&x'=\left(-3x+3\frac{x+\sqrt{2/3}}{3+2\omega_\textsc{bd}}\,\xi\right)\left(1+\sqrt{6}x-\omega_\textsc{bd}x^2\right),\nonumber\\
&&\xi'=-\sqrt{6}x\left[k+m-km-1\right.\nonumber\\
&&\left.\;\;\;\;\;\;\;\;\;\;\;\;\;\;\;\;\;\;\;\;\;\;\;\;\;\;\;\;\;\;\;\;\;\;\;\;\;-(k+m-2)\,\xi-\xi^2\right].\label{ode-vac-comb-exp}\eea 

In this case (assuming that $m>k$), since 

\bea \xi=\frac{1-k+(1-m)\left(\frac{N}{M}\right)^2 e^{(m-k)\vphi}}{1+\left(\frac{N}{M}\right)^2 e^{(m-k)\vphi}},\label{xi-comb-exp}\eea as $\vphi$ undergoes $-\infty<\vphi<\infty$ $\Rightarrow\;1-m\leq\xi\leq 1-k$. Hence, the phase space where to look for equilibrium points of the dynamical system (\ref{ode-vac-comb-exp}), is the bounded compact region of the phase plane $(x,\xi)$, given by $$\Psi^\text{c.exp}_\text{vac}=\left\{(x,\xi):\alpha_-\leq x\leq\alpha_+,\;1-m\leq\xi\leq 1-k\right\},$$ where, we recall, $\alpha_\pm=\sqrt{3/2}(1\pm\sqrt{1+2\omega_\textsc{bd}/3})/\omega_\textsc{bd}$ (see Eq. (\ref{x-bound})).

In the case of the cosh and sinh-like potentials, Eq. (\ref{cosh-like}) and (\ref{sinh-like}) respectively, one has:

\bea &&x'=\left(-3x+3\frac{x+\sqrt{2/3}}{3+2\omega_\textsc{bd}}\,\xi\right)\left(1+\sqrt{6}x-\omega_\textsc{bd}x^2\right),\nonumber\\
&&\xi'=-\frac{\sqrt{6}}{k}\,x\left(k^2\mu^2-1+2\xi-\xi^2\right).\label{ode-vac-cosh-sinh-like}\eea 

The difference between the cosh and the sinh-like potentials is in the phase space where to look for critical points of (\ref{ode-vac-cosh-sinh-like}). For the cosh-like potentials one has that the phase space is the following bounded and compact region of the phase plane $$\Psi^\text{cosh}_\text{vac}=\left\{(x,\xi):\alpha_-\leq x\leq\alpha_+,\;1-k\mu\leq\xi\leq 1+k\mu\right\},$$ while, for the sinh-like potentials the phase space is the unbounded region $\Psi^\text{sinh}_\text{vac}=\Psi^\text{sinh-}_\text{vac}\cup\Psi^\text{sinh+}_\text{vac}$, where

\bea &&\Psi^\text{sinh-}_\text{vac}=\left\{(x,\xi):\alpha_-\leq x\leq\alpha_+,\;1+k\mu\leq\xi<\infty\right\},\nonumber\\
&&\Psi^\text{sinh+}_\text{vac}=\left\{(x,\xi):\alpha_-\leq x\leq\alpha_+,\;-\infty<\xi\leq 1-k\mu\right\}.\nonumber\eea 

A distinctive feature of the dynamical systems (\ref{ode-vac-comb-exp}) and (\ref{ode-vac-cosh-sinh-like}), is that the GR--de Sitter critical point with $x=\xi=0$, $$P_\text{dS}:\left(0,0\right)\;\Rightarrow\;H=H_0,\;\vphi=\vphi_0,$$ is shared by all of them. However, as it will be shown in section \ref{problem-sec}, this does not mean that for potentials of the kinds (\ref{comb-exp}), (\ref{cosh-like}), and (\ref{sinh-like}), with arbitrary free parameters, the $\Lambda$CDM model is an equilibrium point of the corresponding dynamical system. As a matter of fact, only for those arrangements of the free parameters which allow that the given potential has the exponential $U\propto\exp\vphi$ as an asymptote, the $\Lambda$CDM model is an equilibrium configuration of the corresponding dynamical system (see the discussion in section \ref{problem-sec}).


\begin{figure*}[t!]\begin{center}
\includegraphics[width=4cm]{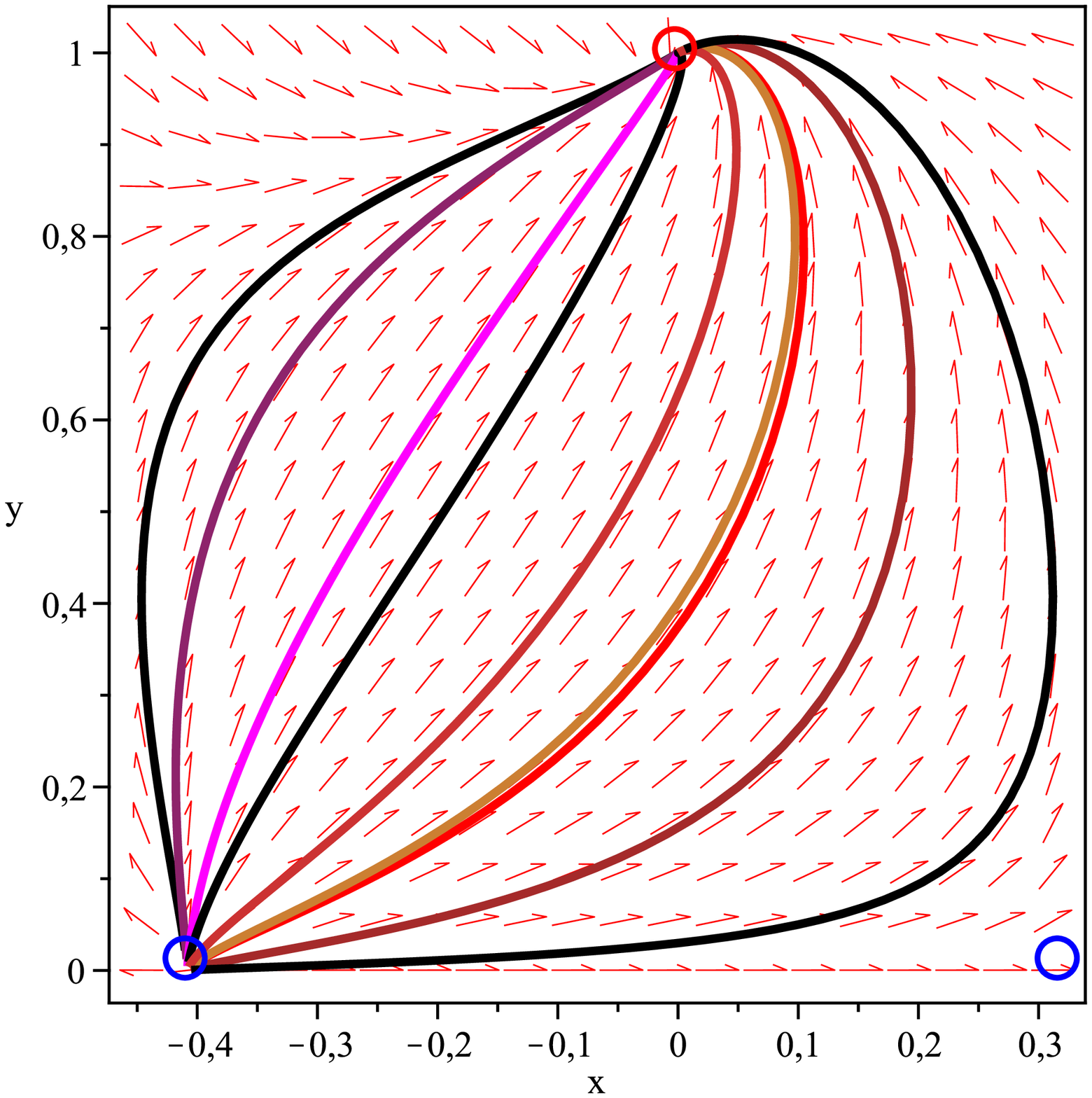}
\includegraphics[width=4cm]{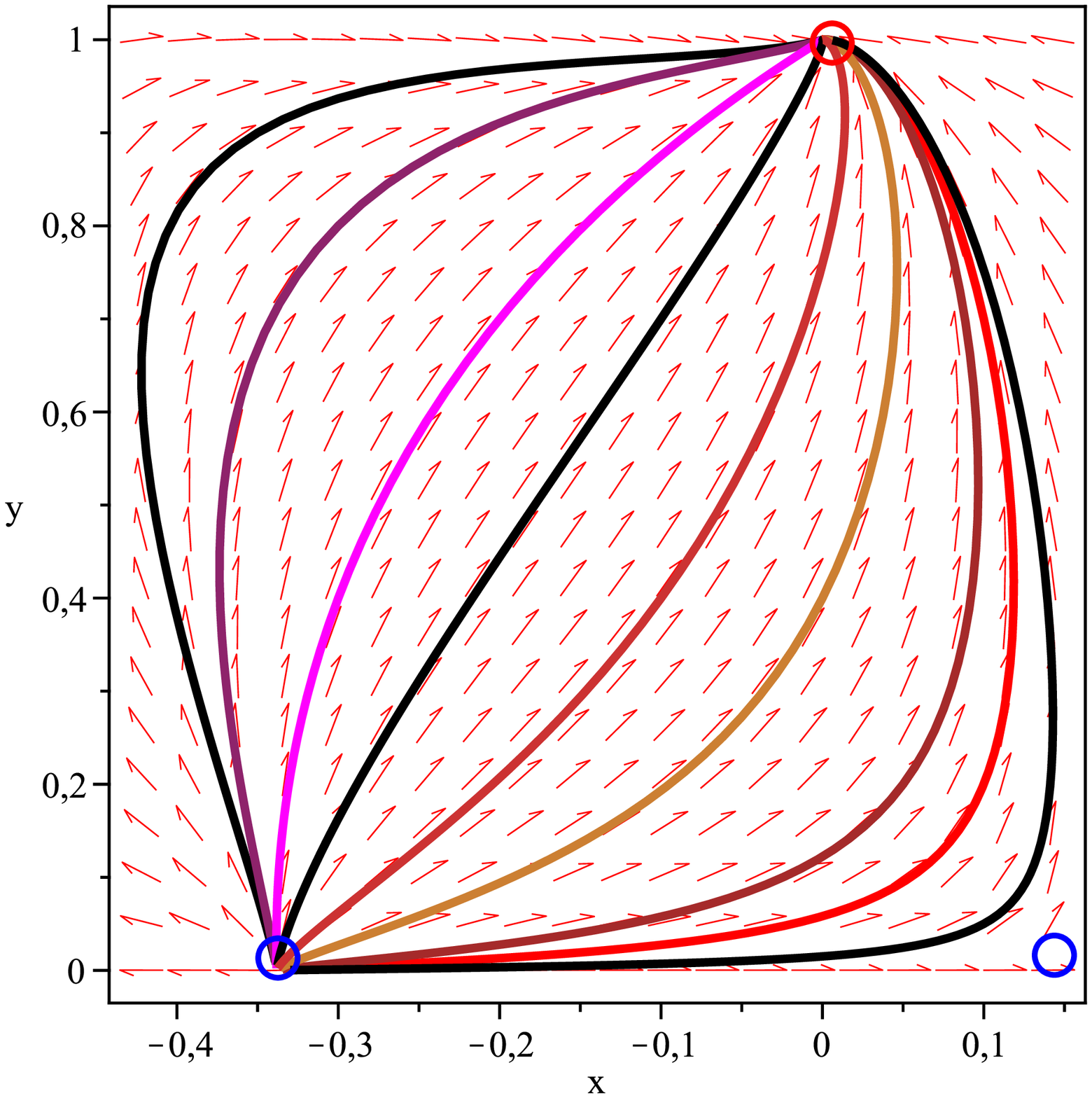}
\includegraphics[width=4cm]{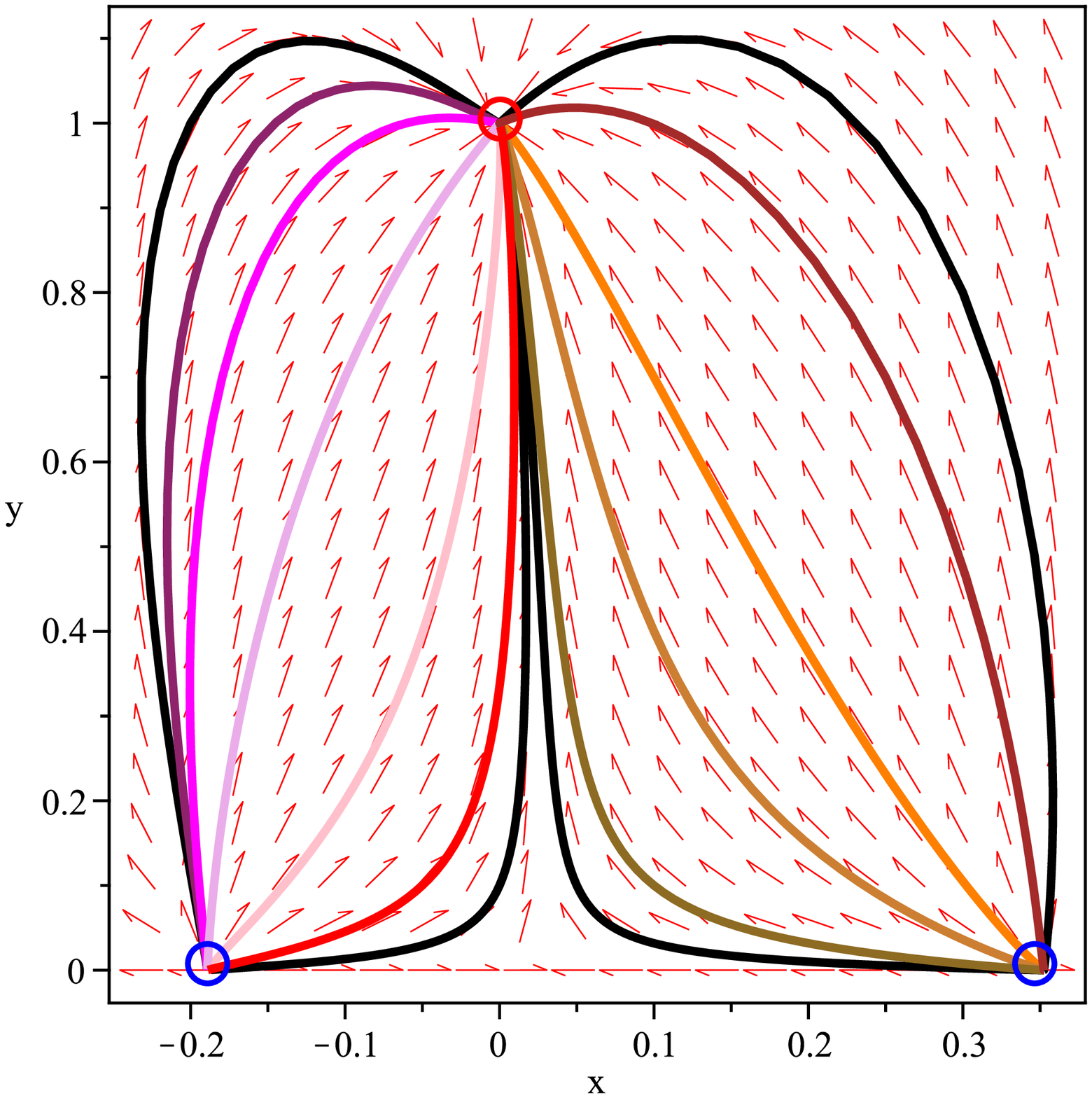}
\includegraphics[width=4cm]{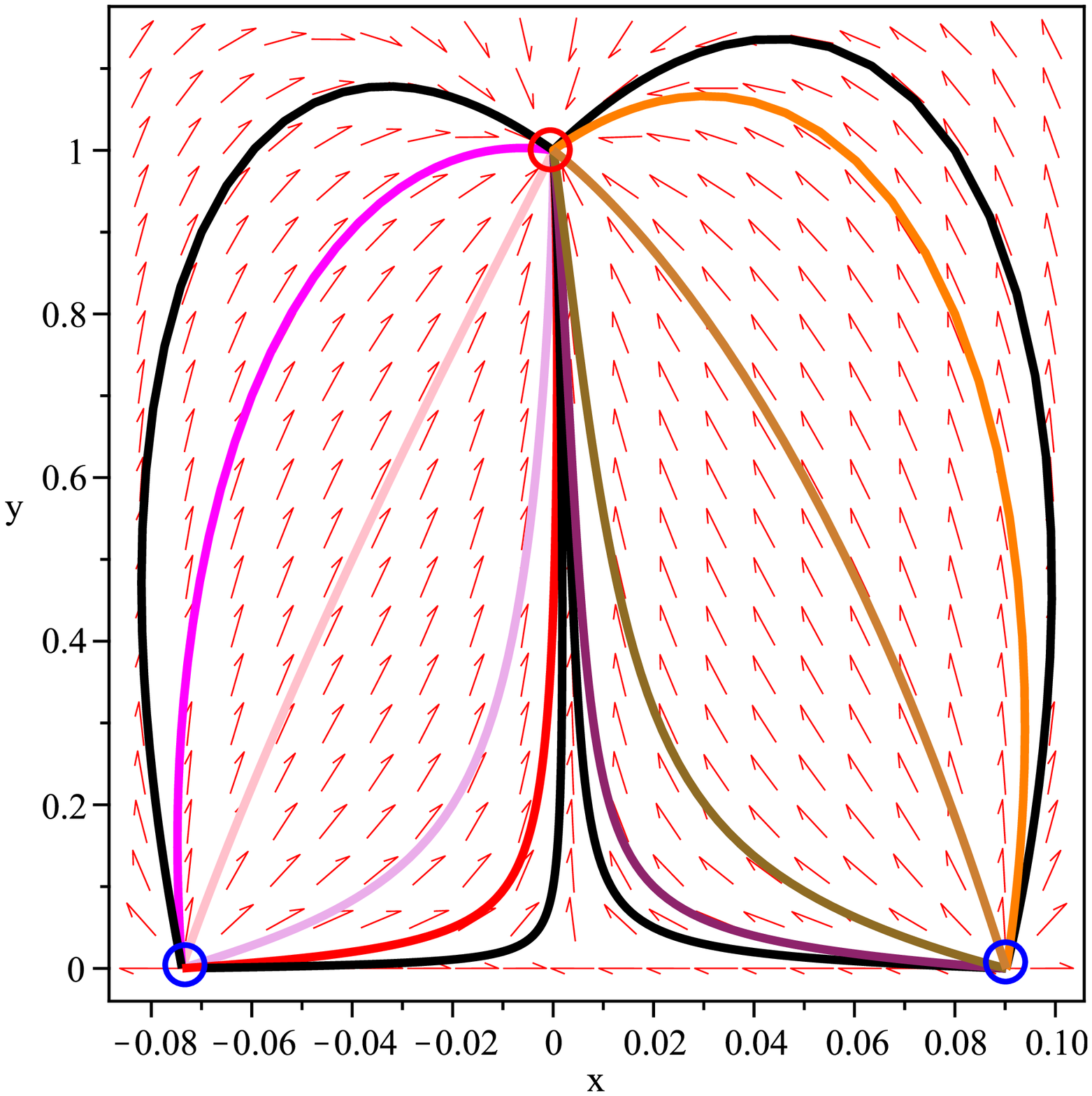}\vspace{0.7cm}
\caption{Phase portrait of the plane-autonomous system of ODE-s (\ref{asode-m}) for $k=1$, corresponding to the exponential potential $U(\vphi)\propto\exp\vphi$, for different values of the BD coupling constant (from left to right): $\omega_\textsc{bd}=0$, $\omega_\textsc{bd}=1.5$, $\omega_\textsc{bd}=15$ and $\omega_\textsc{bd}=150$, respectively. The GR--de Sitter point ($x=0$, $y=1$) is always the future attractor. The stiff-matter solution in the bottom-right corner in each figure is a saddle critical point in the first two cases, while it is a past attractor in the last two cases. The remaining stiff-matter point -- the one in the bottom-left corner -- is always a past attractor.}\label{fig}\end{center}\end{figure*}



\section{Brans--Dicke cosmology with matter}\label{matter-sec}

In the former section we have investigated the dynamical properties of the vacuum Brans--Dicke cosmology in the phase space. Here we shall explore the case when the field equations are sourced by CDM, i. e., by pressureless dust with $w_m=0$, and for exponential potentials (\ref{exp}) only, since, in this latter case, $\xi=1-k$, is a constant. This means that the relevant phase space will be a region of the phase plane $(x,y)$. For this case the autonomous system of ODE-s (\ref{asode}) results in the following plane-autonomous system (see equations (\ref{x-ode-1}) and (\ref{y-ode-1})):

\bea &&x'=-3x\left(1+\sqrt{6}x-\omega_\textsc{bd}x^2\right)\nonumber\\
&&\;\;\;\;\;\;\;\;\;\;\;\;\;\;\;\;\;\;\;\;\;\;\;\;+\frac{3(1-k)}{3+2\omega_\textsc{bd}}\left(x+\sqrt{2/3}\right)y^2\nonumber\\
&&\;\;\;\;\;\;\;\;\;\;\;\;\;\;\;\;\;\;\;\;\;\;\;\;+\frac{1+\sqrt{6}\left(2+\omega_\textsc{bd}\right)\,x}{\sqrt{6}\left(3+2\omega_\textsc{bd}\right)}\,3\Omega^\text{eff}_m,\nonumber\\
&&y'=y\left[3x\left(\omega_\textsc{bd}x-\frac{4-k}{\sqrt 6}\right)+\frac{3(1-k)}{3+2\omega_\textsc{bd}}\,y^2\right.\nonumber\\
&&\left.\;\;\;\;\;\;\;\;\;\;\;\;\;\;\;\;\;\;\;\;\;\;\;\;\;\;\;\;\;\;\;\;\;\;\;\;\;\;\;\;\;+\frac{2+\omega_\textsc{bd}}{3+2\omega_\textsc{bd}}\,3\Omega^\text{eff}_m\right],\label{asode-m}\eea which has physically meaningful equilibrium configurations only within the phase plane: 

\bea &&\Psi_\text{mat}=\left\{(x,y):\;\alpha_-\leq x\leq\alpha_+,\right.\nonumber\\
&&\left.\;\;\;\;\;\;\;\;\;\;\;\;\;\;\;\;\;\;\;\;\;\;\;0\leq y\leq\sqrt{1+\sqrt{6}x-\omega_\textsc{bd}x^2}\right\},\nonumber\eea where we have considered the facts that $\Omega^\text{eff}_m\geq 0$ and $y\in R^+\cup 0$. The critical points of this dynamical system are:

\bea &&P_\text{stiff}:\left(\frac{1-\sqrt{1+2\omega_\textsc{bd}/3}}{\sqrt{2/3}\omega_\textsc{bd}},0\right)\;\Rightarrow\;\Omega^\text{eff}_m=0;\nonumber\\
&&P'_\text{stiff}:\left(\frac{1+\sqrt{1+2\omega_\textsc{bd}/3}}{\sqrt{2/3}\omega_\textsc{bd}},0\right)\;\Rightarrow\;\Omega^\text{eff}_m=0;\nonumber\\
&&P_\text{sc}:\left(\frac{1}{\sqrt{6}(1+\omega_\textsc{bd})},0\right)\;\Rightarrow\nonumber\\
&&\;\;\;\;\;\;\;\;\;\;\;\;\;\;\;\;\;\;\;\;\;\;\;\;\;\;\;\;\;\Omega^\text{eff}_m=\frac{12+17\omega_\textsc{bd}+6\omega^2_\textsc{bd}}{6(1+\omega_\textsc{bd})^2};\nonumber\\
&&P'_\text{sc}:\left(-\frac{\sqrt{3/2}}{k+1},\frac{\sqrt{k+4+3\omega_\textsc{bd}}}{\sqrt{2}(k+1)}\right)\;\Rightarrow\nonumber\\
&&\;\;\;\;\;\;\;\;\;\;\;\;\;\;\;\;\;\;\;\;\;\;\;\;\;\;\;\;\;\Omega^\text{eff}_m=\frac{2k^2-3k-8-6\omega_\textsc{bd}}{2(k+1)^2};\nonumber\\
&&P_*:\left(-\frac{\sqrt{2/3}(k-1)}{k+2+2\omega_\textsc{bd}},\frac{\beta}{k+2+2\omega_\textsc{bd}}\right)\;\Rightarrow\nonumber\\
&&\;\;\;\;\Omega^\text{eff}_m=\frac{12-6k-6k^2+\left(7-2k-5k^2\right)\omega_\textsc{bd}}{2(k+2+2\omega_\textsc{bd})^2},\label{c-points}\eea where, in the last critical point we have defined the parameter: $\beta=\sqrt{1+2\omega_\textsc{bd}/3}\sqrt{8+6\omega_\textsc{bd}-k(k-2)}$. The equilibrium points $P_\text{stiff}$ and $P'_\text{stiff}$ represent stiff-fluid solutions, meanwhile the remaining points represent scaling between the energy density of the dilaton and the CDM. 

Let us to focus into two of the above critical points: $P'_\text{sc}$ and $P_*$. As it was for vacuum BD cosmology, the de Sitter critical point does not arise unless $k=1$. In this latter case ($k=1$), for the last equilibrium point in Eq. (\ref{c-points}), one gets: $$P_*:\left(0,1\right),\;q=-1\;(H=H_0),\;\Omega^\text{eff}_m=0,\;\lambda_{1,2}=-3,$$ where $\lambda_1$ and $\lambda_2$ are the eigenvalues of the linearization matrix around $P_*:(0,1)$. This means that, for the exponential potential $U(\vphi)\propto\exp\vphi$, the GR--de Sitter solution is an attractor of the dynamical system (\ref{asode-m}). This is evident from the FIG. \ref{fig}, where the attractor character of the critical point $(0,1)$ is apparent. 

For the scaling point $P'_\text{sc}$, the deceleration parameter is given by $$q=\frac{k-2}{2(k+1)},$$ so that, for $k=0$, which corresponds to the constant potential $U=U_0$, the BD--de Sitter solution is obtained $$q=-1\;\Rightarrow\;a(t)\propto e^{H_0 t},\;e^{-\vphi}\rho_m=const.$$ However, since $$\Omega_m=\frac{2k^2-3k-8-6\omega_\textsc{bd}}{2(k+1)^2},$$ at $k=0$, $\Omega^\text{eff}_m=-(4+3\omega_\textsc{bd})$, is a negative quantity, unless the Brans--Dicke coupling parameter falls into the very narrow interval $-3/2<\omega_\textsc{bd}\leq-4/3$. Hence, for $k=0$, but for $-1.5<\omega_\textsc{bd}\leq-1.33$, the point $P'_\text{sc}$ does not actually belong in the phase space $\Psi_\text{mat}$.


\section{(Non)emergence of the $\Lambda$CDM phase from the Brans--Dicke cosmology}\label{problem-sec}

This problem has been generously discussed before in the reference \cite{hrycyna}. The conclusion on the emergence of the $\Lambda$CDM cosmology starting from the Brans--Dicke theory, seems to be supported by the existence of a de Sitter phase, which was claimed to be independent on the concrete form of the self-interaction potential of the dilaton field in \cite{hrycyna, hrycyna-1}, and then, in Ref. \cite{hrycyna-2} the same authors somewhat corrected their previous claim. In this section we shall address this problem and we will clearly show that, in general (but for the exponential potential $U(\vphi)\propto e^\vphi$), the $\Lambda$CDM model is not an attractor of the FRW--BD cosmology. 

Before we go any further, we want to make clear that the latter statement on the non-universality of the GR--de Sitter equilibrium point, does not forbids the possible existence of exact de Sitter solutions for several choices of the self-interaction potential (see, for instance, Ref. \cite{odintsov-ref}). What the statement means is that, in case such solutions are found, these would not be generic solutions, but very particular (unstable) solutions instead, which are unable to represent any sensible cosmological scenario. 

Before we start our discussion, it will be useful to state that a de Sitter solution arises whenever $$q=-1\;\Rightarrow\;\dot H=0\;\Rightarrow\;H=H_0\;\Rightarrow\;a(t)\propto e^{H_0 t}.$$ This condition can be achieved even if $x\neq 0$. However, only when $$x=0\;\Rightarrow\;\dot\vphi=0\;\Rightarrow\;\vphi=\vphi_0,$$ the de Sitter solution can lead to the $\Lambda$CDM model, where by $\Lambda$CDM model we understand the FRW cosmology within the frame of Einstein's GR, with a cosmological constant $\Lambda$ and cold dark matter as the sources of gravity. Actually, only if $\vphi=\vphi_0$, is a constant, the action (\ref{dbd-action}) -- up to a meaningless factor of $1/2$ -- is transformed into the Einstein-Hilbert action plus a matter source: $$S=\frac{1}{8\pi G_N}\int d^4x\sqrt{|g|}\left\{R-2U_0\right\}+2\int d^4x\sqrt{|g|}{\cal L}_m,$$ where $e^{\vphi_0}=1/8\pi G_N$. When ${\cal L}_m$ is the Lagrangian of CDM, the latter action -- compare with Eq. (\ref{lcdm}) -- is the mathematical expression of what we call as the $\Lambda$CDM cosmological model. In the remaining part of this section we shall discuss on the (non)universality of the $\Lambda$CDM equilibrium point. For this purpose, in order to find useful clues, we shall explore first the simpler situation of vacuum BD cosmology and, then, the Brans--Dicke cosmology with CDM will be explored.

\subsection{Vacuum FRW--BD cosmology}\label{w-bound}

In this simpler situation the de Sitter phase arises only if assume an exponential potential of the form $$U(\vphi)\propto\exp\vphi\;\Rightarrow\;V(\phi)=M^2\phi^2,$$ which means that $\xi=0$ and $\Gamma=1$, are both completely specified, or if $\xi=1$, i. e., if $$U(\vphi)=M^2\;\Rightarrow\;V(\phi)=M^2\phi.$$ As a matter of fact, as shown in section \ref{vacuum-sec}, for exponential potentials of the general form: $$U(\vphi)=M^2\ e^{k\vphi}\;\Rightarrow\;V(\phi)=M^2\phi^{k+1},$$ with $k\neq 1$ and $k\neq 0$, the de Sitter critical point does not exist. In other words, speaking in terms of the original BD variables: but for the quadratic and the lineal monomials, $V(\phi)\propto\phi^2$ and $V(\phi)\propto\phi$, respectively -- also for those potentials which asymptote to either $\phi^2$ or $\phi$ -- the de Sitter solution is not an equilibrium point of the corresponding dynamical system. 

Now we want to show that, even when a de Sitter solution is a critical point of (\ref{x-ode-vac-cc}), the existence of a de Sitter equilibrium point in the vacuum BD cosmology, by itself, does not warrant that the $\Lambda$CDM model is approached. As an illustration of this statement, let us choose the vacuum FRW--BD cosmology driven by a constant potential (see subsection \ref{cc}). In this case one of the equilibrium points of the dynamical system (\ref{x-ode-vac-cc}): $$x_1=1/\sqrt{6}(1+\omega_\textsc{bd})\neq 0,$$ corresponds to the de Sitter solution since $$q=-1\;\Rightarrow\;\frac{\dot H}{H^2}=0\;\Rightarrow\;H=H_0.$$ The tricky situation here is that, although the de Sitter solution ($H=H_0$) is a critical point of the dynamical system (\ref{x-ode-vac-cc}), the $\Lambda$CDM model is not mimicked. Actually, at $x_1$, 

\bea &&x=\frac{\dot\vphi}{\sqrt{6}\,H}=\frac{1}{\sqrt{6}(1+\omega_\textsc{bd})}\;\Rightarrow\nonumber\\
&&\dot\vphi=\frac{H_0}{1+\omega_\textsc{bd}}\;\Rightarrow\;\vphi(t)=\frac{H_0\,t}{1+\omega_\textsc{bd}}+\vphi_0,\nonumber\eea i. e., the scalar field evolves linearly with the cosmic time $t$. This point corresponds to BD theory and not to GR since, while in the latter the Newton's constant $G_N$ is a true constant, in the former the effective gravitational coupling (the one measured in Cavendish-like experiments) evolves with the cosmic time: $$G_\text{eff}=\frac{4+2\omega_\textsc{bd}}{3+2\omega_\textsc{bd}}\,e^{-\vphi}\;\Rightarrow\;\frac{\dot G_\text{eff}}{G_\text{eff}}=-\frac{H_0}{1+\omega_\textsc{bd}}.$$ Taking the Hubble time to be $t_0=13.817\times 10^9$ yr (as, for instance, in \cite{hrycyna}), i. e., the present value of the Hubble constant $H_0=7.24\times 10^{-11}$ yr$^{-1}$, one gets 

\bea \frac{\dot G_\text{eff}}{G_\text{eff}}=-\frac{1}{1+\omega_\textsc{bd}}\,7.24\times 10^{-11}\,\text{yr}^{-1}.\label{estimate}\eea 

As a consequence of the above, if consider cosmological constraints on the variability of the gravitational constant \cite{uzan-rev}, for instance the ones in \cite{cosmo}, which uses WMAP-5yr data combined with SDSS power spectrum data: $$-1.75\times 10^{-12}\,\text{yr}^{-1}<\frac{\dot G}{G}<1.05\times 10^{-12}\,\text{yr}^{-1},$$ or the ones derived in Ref. \cite{cosmo-1}, where the dependence of the abundances of the D, $^3$He, $^4$He, and $^7$Li upon the variation of $G$ was analyzed: $$|\dot G/G|<9\times 10^{-13}\,\text{yr}^{-1},$$ from Eq. (\ref{estimate}) one obtains the following bounds on the value of the BD coupling constant: $$\omega_\textsc{bd}>40.37\;|\;\omega_\textsc{bd}<-69.95,$$ and $$\omega_\textsc{bd}>79.44\;|\;\omega_\textsc{bd}<-81.44,$$ respectively. These constraints contradict the results of \cite{hrycyna, hrycyna-1}, and are more in the spirit of the estimates of \cite{bd-coupling, aquaviva} (see, also, Ref. \cite{chiva}).

\subsection{Other potentials}

As seen in section \ref{other}, for other potentials, such as the combination of exponentials (\ref{comb-exp}), the cosh (\ref{cosh-like}) and sinh-like (\ref{sinh-like}) potentials, the GR--de Sitter solution is a critical point of the corresponding dynamical system. However, do not get confused: the above statement is not true for any arrangement of the free constants. 

Take, for instance, the combination of exponentials. The GR--de Sitter point $x=\xi=0$ entails that (see Eq. (\ref{xi-comb-exp})), either $k=m=1$ $\Rightarrow\;\xi=0$, or, for $m=1$, arbitrary $k$, the point is asymptotically approached as $\vphi\rightarrow\infty$ if $k<1$. In the former case ($k=m=1$) the combination of exponentials $$U(\vphi)=M^2 e^{k\vphi}+N^2 e^{m\vphi},$$ coincides with the simple exponential (\ref{exp}), $U(\vphi)=(M^2+N^2)\,e^\vphi$, while in the latter case ($m=1$, $k$ arbitrary), assuming that $k<1$, the above potential tends asymptotically ($\vphi\rightarrow\infty$) to the exponential $U(\vphi)\approx N^2 e^\vphi$. 

For the cosh and sinh-like potentials one has (see section \ref{sub-cosh-sinh}): 

\bea U(\vphi)=M^2\left(e^{\mu\vphi}\pm e^{-\mu\vphi}\right)^k,\label{cosh-sinh}\eea where the ``$+$'' sign is for the cosh potential, while the ``$-$'' sign is for the sinh potential, and the $2^{-k}$ has been absorbed in the constant factor $M^2$. On the other hand, one has the following relationships (see section \ref{sub-cosh-sinh}): $$\xi=1-k\mu\frac{e^{\mu\vphi}-e^{-\mu\vphi}}{e^{\mu\vphi}+e^{-\mu\vphi}},\;\xi=1-k\mu\frac{e^{\mu\vphi}+e^{-\mu\vphi}}{e^{\mu\vphi}-e^{-\mu\vphi}},$$ where the left-hand equation is for the cosh-like potential, while the right-hand one is for the sinh-like potential. Since at the GR--de Sitter point: $x=\xi=0$, then, from the above equations it follows that this critical point exists, for the cosh and sinh-like potentials, only if $k\mu=1$, in which case, the mentioned potentials (\ref{cosh-sinh}) asymptotically approach to the exponential as $\vphi\rightarrow\infty$: $$U(\vphi)\approx M^2\,e^{k\mu\vphi}=M^2\,e^\vphi.$$

Summarizing: Only for the exponential potential $U(\vphi)\propto\exp\vphi$, or for any other potential which, as $\vphi\rightarrow\infty$, tends asymptotically to the exponential $\exp\vphi$, the GR--de Sitter solution is an attractor of the dynamical system (\ref{x-xi-ode-vac}). This is easily visualized if realize that, by the definition of the variable $\xi$: $$\xi=1-\frac{\der_\vphi U}{U}.$$ Hence, if assume $\xi=0$, which is a necessary condition for the existence of the GR--de Sitter point, then, necessarily: $$\frac{\der_\vphi U}{U}=1\;\Rightarrow\;U(\vphi)\propto e^\vphi.$$

\subsection{FRW--BD cosmology with matter}

In the case when we consider a matter source for the Brans--Dicke equations of motion, in particular CDM, the existence of a de Sitter critical point with $x=0$ $\Rightarrow\;\dot\vphi=0$ -- which means that the effective gravitational coupling is a real constant that can be made to coincide with the Newton's constant -- is to be associated with the $\Lambda$CDM model. 

The autonomous system of ODE-s that can be obtained out of the cosmological FRW--BD equations of motion, when these are sourced by CDM, is the one in Eq. (\ref{asode-m}). The critical points of this dynamical system are given in Eq. (\ref{c-points}). Notice that only one of them: $$P_*:\left(-\frac{\sqrt{2/3}(k-1)}{k+2+2\omega_\textsc{bd}},\frac{\beta}{k+2+2\omega_\textsc{bd}}\right),$$ where $\beta=\sqrt{1+2\omega_\textsc{bd}/3}\sqrt{8+6\omega_\textsc{bd}-k(k-2)}$, can be associated with GR--de Sitter expansion, i. e., with what we know as the $\Lambda$CDM model, in the special case when $k=1$. In this latter case $P_*:(0,1)$. Since we are considering exponential potentials of the form in Eq. (\ref{exp}), then, the GR--de Sitter equilibrium configuration is associated, exclusively, with the potential $$\frac{\der_\vphi U}{U}=k=1\;\Rightarrow\;U(\vphi)\propto e^\vphi.$$

Although in section \ref{matter-sec} we have considered only exponential potentials in FRW--BD cosmology with background dust, it is clear that the result remains the same as for the vacuum case: Only for the exponential potential $U(\vphi)\propto\exp\vphi$, or for potentials that approach asymptotically to $\exp\vphi$, the GR--de Sitter solution is an equilibrium configuration of the corresponding dynamical system.


\section{discussion}\label{discu-sec}

Why do our results differ from those in Ref. \cite{hrycyna, hrycyna-1}, even when the tools used are the same? Although we should not aim here at a detailed analysis of the work of \cite{hrycyna, hrycyna-1}, nevertheless we can guess what is going on. To start with we shall concentrate, specifically, in the result related with what the authors of \cite{hrycyna, hrycyna-1} call as the asymptotic value of the scalar field mass at the de Sitter point, which is the value of the BD scalar field mass computed with the help of the following known equation \cite{hrycyna, hrycyna-1, mass}: $$m^2=\frac{2}{3+2\omega_\textsc{bd}}\left[\phi \der^2_\phi V(\phi)-\der_\phi V(\phi)\right],$$ or, in terms of the field variables $\vphi$ and $U=U(\vphi)$ in Eq. (\ref{vphi}), the mass squared of the dilaton:

\bea m^2=\frac{2}{3+2\omega_\textsc{bd}}\left(\der^2_\vphi U-U\right),\label{mass-f}\eea evaluated at the GR--de Sitter equilibrium point. 

According to  \cite{hrycyna, hrycyna-1}, the asymptotic value of the scalar field mass $m|_*$, at the de Sitter point, is given by $$m|_*\approx\frac{1.84\times 10^{-33}}{\sqrt{3+2\omega_\textsc{bd}}}\;\text{eV}.$$ Then the authors constrain the BD coupling parameter $\omega_\textsc{bd}$ by contrasting the above value $m|_*$ with known estimates signaling at $m|_*\sim 10^{-22}$ eV. The obtained bound $$\omega_\textsc{bd}\approx -\frac{3}{2}+10^{-22},$$ coincides with the conformal coupling value of $\omega_\textsc{bd}$. This value of the BD coupling parameter is a singular value and, if matter is taken into account, is very problematic since, consistency of the BD motion equations require that only traceless matter can be coupled to the BD scalar field if $\omega_\textsc{bd}=-3/2$. The above bound on $\omega_\textsc{bd}$ is to be contrasted with our result in the subsection \ref{w-bound}, which was based on the analysis of the normalized ratio of variation of the gravitational coupling $\dot G/G$, which clearly excluded the possibility of $\omega_\textsc{bd}=-3/2$. Then, what is going on?

Let us start to develop our reasoning line by recalling that, as properly noted in \cite{hrycyna, hrycyna-1}, the mass squared of the dilaton (\ref{mass-f}) can be written as a function of the phase space variables:

\bea m^2=m^2(x,y,\xi)=\frac{6H^2 y^2}{3+2\omega_\textsc{bd}}\left[\left(1-\xi\right)^2\Gamma(\xi)-1\right].\label{mass-y-xi}\eea If consider, for instance, the vacuum BD cosmology -- see section \ref{vacuum-sec} -- one has that: $$m^2=\frac{6H^2}{3+2\omega_\textsc{bd}}\left(1+\sqrt{6}x-\omega_\textsc{bd}x^2\right)\left[(1-\xi)^2\Gamma-1\right].$$ Worth noticing that, since the phase space is bounded by the condition $\alpha_-\leq x\leq\alpha_+$, where the $\alpha_\pm$ in Eq. (\ref{x-bound}) are the roots of the second order algebraic equation $1+\sqrt{6}x-\omega_\textsc{bd}x^2=0$, then the mass squared of the dilaton is a non-negative quantity, provided that $\Gamma\geq 1/(1-\xi)^2$. One immediately notices that, at the stiff-dilaton (vacuum) solutions, where $\Omega_K^\text{eff}=1$, since $\Omega_K^\text{eff}=\omega_\textsc{bd}x^2-\sqrt{6}x$, then the mass squared of the dilaton vanishes.\footnote{If introduce the dilaton's dimensionless mass squared density $$\Omega_{m^2}=\frac{m^2}{3H^2}=\frac{2(k^2-1)}{3+2\omega_\textsc{bd}}\left(1+\sqrt{6}x-\omega_\textsc{bd}x^2\right),$$ it is seen that it vanishes at the boundaries $x=\alpha_\pm$, and it is a maximum at $x=\sqrt{3/2}/\omega_\textsc{bd}$, where $\Omega_{m^2}^\text{max}=(k^2-1)/\omega_\textsc{bd}$. The quantity $\Omega_{m^2}$ is zero for the exponential potentials $U_\pm\propto\exp\pm\vphi$, for which $k=\pm 1$.} 

For the exponential potential $U(\vphi)=M^2\exp(k\vphi)$, for instance, since $\xi=1-k$, and $\Gamma=1$, then $$m^2=\frac{6H^2(k^2-1)}{3+2\omega_\textsc{bd}}\left(1+\sqrt{6}x-\omega_\textsc{bd}x^2\right).$$ In order for the dilaton mass squared to be non-negative, it is required that $k^2\geq 1$. We see that, for the particular case $k=1$, i. e., for the specific exponential potential $U(\vphi)=M^2\exp\vphi$, the dilaton is a massless degree of freedom. Hence, at the GR--de Sitter equilibrium configuration the mass of the dilaton is necessarily vanishing. 

A general demonstration of the above statement can be based in Eq. (\ref{mass-f}), where no particular considerations on the potential (neither on the matter content of the BD theory) are made. As a matter of fact Eq. (\ref{mass-f}) is the adopted definition of the mass squared of the dilaton in the Jordan frame \cite{mass}, which is the frame considered in this paper. One sees that the mass squared of the dilaton $m^2\propto\der_\vphi^2 U-U$, vanishes provided that $$\der_\vphi^2 U=U\;\Rightarrow\;U(\vphi)\propto e^{-\vphi}\;|\;U(\vphi)\propto e^\vphi,$$ where, as shown above, for the (growing) exponential potential $U\propto\exp\vphi$, the GR--de Sitter critical point is an attractor of the corresponding dynamical system. 

Given that the scalar field is necessarily massless at the GR--de Sitter point, which would be the meaning of the tiny, yet non-vanishing, asymptotic value $m|_*$ computed in \cite{hrycyna}? In this regard notice that the computations in \cite{hrycyna, hrycyna-1} are based on the linearized solutions (perturbations would be more precise) around the de Sitter point, which are valid up to linear terms in the initial conditions. Besides, in order to obtain the bound $\omega_\textsc{bd}\approx -3/2+10^{-22}$ on the BD coupling parameter, the authors of \cite{hrycyna, hrycyna-1} assumed what they called as ``special initial conditions''. Then, the mass of the BD scalar field computed in the mentioned references, is the mass of the field at the linearized (perturbed) solutions around the de Sitter point, but not at the point itself, where the dilaton is actually massless, as we have shown above. 

The next question would be: which is the actual meaning of the linearized solutions? The linearized solutions correspond to points in the phase space which are very close to the stable equilibrium point -- the de Sitter critical point in the present case -- so that the linear approximation takes place: 

\bea x(\tau)\approx x_c+\epsilon_x(\tau),\;y(\tau)\approx y_c+\epsilon_y(\tau),\label{line-sol}\eea where $x_c$, $y_c$ are the coordinates of the given equilibrium point, and the perturbations $\epsilon_x\sim\epsilon_y\ll 1$, are very small. These solutions can be viewed as small deformations of the stable GR--de Sitter solution. Just as an illustration, let us consider the FRW--BD theory driven by the exponential potential $U(\vphi)\propto\exp\vphi$, in a background of CDM (see section \ref{matter-sec}). The small perturbations around the de Sitter point $P:(0,1)$, very quickly tend to vanish, restoring the system into the stable equilibrium state: $\epsilon_x\sim\epsilon_y\propto\exp(-3\tau)$, where we have taken into account that the eigenvalues of the linearization matrix at $(0,1)$, coincide: $\lambda_1=\lambda_2=-3$. Then, the linearized solutions around the GR--de Sitter point look like:\footnote{Here $A$ and $B$ are integration constants, which depend linearly on the initial conditions $x(0)$, $y(0)$, and on other free parameters such as $\omega_\textsc{bd}$.} 

\bea &&x(\tau)\approx A\,e^{-3\tau}\;\Rightarrow\;\vphi(a)\approx-\sqrt\frac{2}{3}\frac{A}{a^3}+\vphi_0,\nonumber\\
&&y(\tau)\approx 1+B\,e^{-3\tau}\;\Rightarrow\;H^2(a)\approx\frac{M^2\,e^{\vphi_0-\sqrt\frac{2}{3}\frac{A}{a^3}}}{3(1+\frac{B}{a^3})^2}.\nonumber\eea These will eventually (perhaps very quickly) decay into the stable de Sitter solution: 

\bea &&x(\tau)=0\;\Rightarrow\;\vphi=\vphi_0,\nonumber\\
&&y(0)=1\;\Rightarrow\;H=H_0=M\,e^{\vphi_0/2}/\sqrt{3}.\nonumber\eea


\begin{figure}[t!]\begin{center}
\includegraphics[width=4cm]{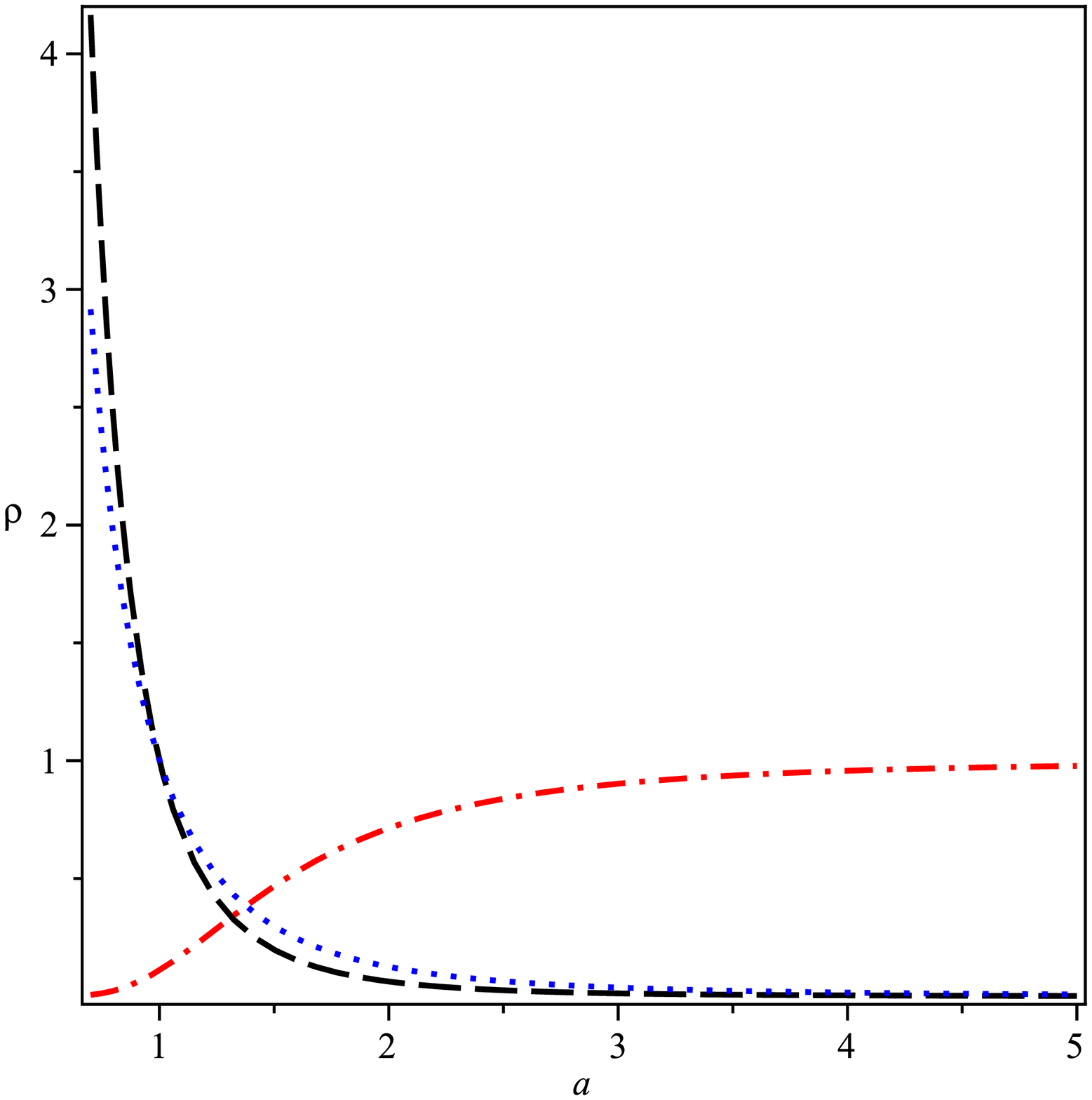}
\includegraphics[width=4cm]{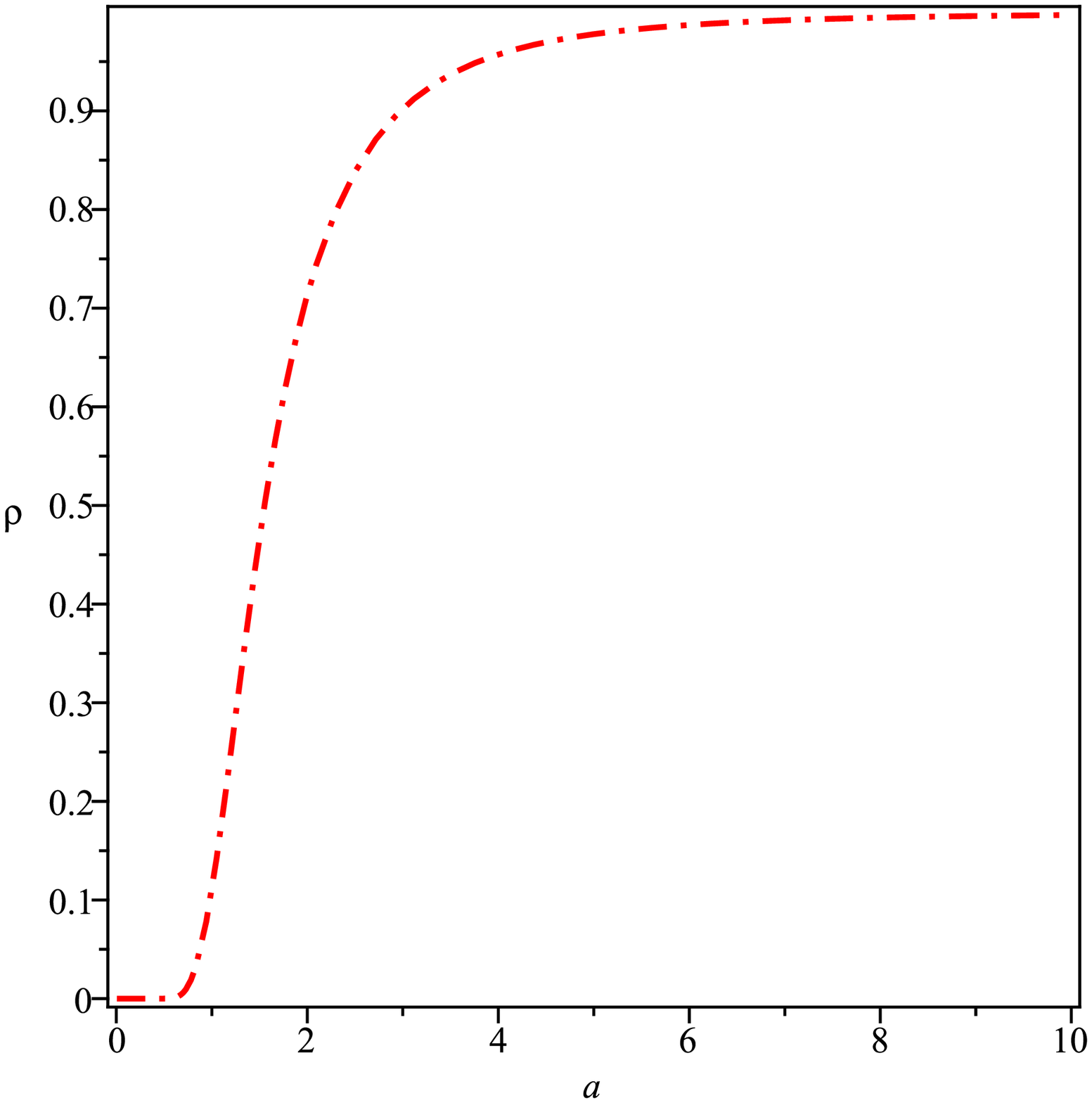}\vspace{0.7cm}
\caption{Plots of the energy density of radiation $\rho_\text{rad}=A^2/a^4$ (dash), dust $\rho_\text{dust}=B^2/a^3$ (dots), and of the perturbed solution $\rho_\text{pert}=M^2 e^{\vphi_0-\sqrt{2/3}A/a^3}(1+B/a^3)^{-2}$ (dash-dots), vs the scale factor $a$, for arbitrarily chosen values of the constants: $A=B=M=1$, $\vphi_0=0$. The plot of the energy density of matter in the perturbed solution is separately shown in the right-hand panel.}\label{fig-density}\end{center}\end{figure}


In the FIG. \ref{fig-density} the plots of the energy density of radiation, dust, and of the ``energy density of the perturbed solution'',  

\bea \rho_\text{rad}=\frac{A^2}{a^4},\;\rho_\text{dust}=\frac{B^2}{a^3},\;\rho_{pert}=\frac{M^2\,e^{\vphi_0-\sqrt\frac{2}{3}\frac{A}{a^3}}}{(1+\frac{B}{a^3})^2},\nonumber\eea respectively, are shown as functions of the scale factor $a$, for arbitrarily chosen values of the free constants. Notice that, while the energy density of ``normal matter'' such as radiation and dust, dilutes with the cosmic expansion, the background energy density in the linearized solution grows with the cosmic time, until, eventually, the perturbed solution decays into the de Sitter expansion with a constant (non-evolving) energy density. This means two things: first, that the background density of the linearized solution behaves as phantom matter during the decay into the stable equilibrium (de Sitter) state, and, second, that the background energy density of the linearized solutions is always smaller than the energy density of the stable de Sitter solution.

We can say that the linearized solutions have a small life-time in the sense that, within a very small amount of ``time'' $\tau$, they decay into the stable solution. It is then clear that the mass of the dilaton computed at linearized solutions, would be highly dependent on the assumed initial conditions, in contrast to the mass of the dilaton at the de Sitter point. Actually, while the mass of the field at perturbed (unstable) solutions, depends on the way the perturbations are generated, at the GR--de Sitter attractor -- being a stable equilibrium configuration -- the field is massless regardless of the initial conditions. Hence, making cosmological predictions on the base of linearized/perturbed solutions (around equilibrium points) is meaningless due to the loss of predictability which is associated with the strong dependence on the initial conditions. The only useful information the dynamical systems theory allow to extract from the given cosmological dynamical system, is encoded in the equilibrium points themselves, but not in the (linear) perturbations around them. The latter serve only as probes to test the stability of the given critical point.

The same reasoning line applies to the computation of other derived quantities such as the ratio $\dot G/G$ (see section \ref{w-bound}). In section \ref{w-bound} we have shown that, at the GR--de Sitter point $$\frac{\dot G_\text{eff}}{G_\text{eff}}=-\frac{H_0}{1+\omega_\textsc{bd}}=-\frac{1}{1+\omega_\textsc{bd}}\,7.24\times 10^{-11}\,\text{yr}^{-1},$$ i. e., the ratio of the variation of the gravitational coupling is a negative quantity (recall that we are considering cosmic expansion exclusively, so that $H=H_0\geq 0$). Hence, contrary to the result of \cite{hrycyna-1}, the gravitational coupling decreases during the cosmic expansion, resulting in the weakening of the strength of gravity along the cosmic history. The explanation of the discrepancy of our result with the corresponding one in Ref. \cite{hrycyna-1}, is similar to the explanation given above to the discrepancy in the bounds on $\omega_\textsc{bd}$. While our computation of the quantity $\dot G/G$ is done at the stable GR--de Sitter equilibrium point, meaning that our result is quite independent of the initial conditions, the corresponding computation in \cite{hrycyna-1} is done at perturbed (linearized) solutions which, as explained above, are unstable and very quickly decay into the stable de Sitter state. The resulting computations are highly dependent on the initial conditions chosen to generate the given perturbation around the de Sitter point. 

We find no reason to believe that we are living in one such perturbed solution and not in the equilibrium configuration itself (the GR--de Sitter critical point). Besides, if one wants to avoid the cosmic coincidence problem, an equilibrium configuration which attracts the cosmic history into a GR--de Sitter stage, is all what one needs. Making definitive conclusions about the entire cosmic history, based in computations made at a perturbed solution (here we are thinking in the conclusion on the weakening of the strength of gravity at early times in Ref. \cite{hrycyna-1}, based in the positivity of $\dot G/G$), is potentially misleading. The positivity of $\dot G/G$ at a perturbed solution, when compared with the negative value of $\dot G/G$ at the stable equilibrium point, may only mean that gravity is a bit weaker at the perturbed solution than it is at the stable critical point, no more. This result is closely connected with the fact that the background energy density of the linearized solutions is always smaller than the energy density of the stable de Sitter solution (see the related discussion above).

In a similar way we want to cast reasonable doubt on the conclusion in Ref. \cite{hrycyna-1} about the correspondence of the mean value of the BD parameter $\omega_\textsc{bd}$ with the coupling parameter between the dilaton and the graviton in the low-energy limit of the string effective theory $\omega_\textsc{bd}=-1$ \cite{wands-rev}. As we have discussed above, in this case, in order to check the observational data, the authors used expressions for the normalized (squared) Hubble function $(H(a)/H(a_0))^2$, which were computed at the linearized solutions (up to linear terms in the initial conditions), but not at the stable GR--de Sitter critical point itself. This means that, as previously discussed, the results of \cite{hrycyna, hrycyna-1} are highly dependent on the initial conditions and, hence, useless to make cosmological predictions.


\section{conclusion}\label{conclu-sec}

In the present paper we have explored the asymptotic properties of FRW--BD cosmological models (\ref{efe}), driven by a variety of self-interaction potentials $U(\vphi)$. For this purpose we have used the simplest tools of the dynamical systems theory \cite{hrycyna, hrycyna-1, hrycyna-2, classic-books, wands, coley, amendola, copeland-rev, copeland, luis, bohmer-rev, epj-2015, genly, genly-1, cosmology-books}. We have shown that, in spite of known results of previously published work \cite{hrycyna, hrycyna-1, hrycyna-2}, the GR--de Sitter phase is not an universal attractor of the BD theory. Only for the specific exponential potential $U(\vphi)\propto\exp\vphi$, which, in terms of the original BD field $\phi$, amounts to the quadratic monomial $V(\phi)\propto \phi^2$, or for potentials which asymptotically approach to $\exp\vphi$ ($\phi^2$), the GR--de Sitter phase is a stable critical point, i. e., a future attractor in the phase space. We have shown, also, that at the GR--de Sitter critical point, as well as at the stiff-matter equilibrium configurations, the effective mass of the dilaton $m^2$ in Eq. (\ref{mass-f}), vanishes.

We have learned that physically meaningful conclusions can be based only on computations performed at the equilibrium configurations as, for instance, at the stable GR--de Sitter critical point. On the contrary, the results based on computations made at perturbed solutions are highly dependent on the initial conditions chosen and, hence, useless to make physically meaningful predictions. 

In particular, the computations performed at the GR--de sitter critical point yield to bounds on the value of the BD coupling parameter $\omega_\textsc{bd}>40.37\;|\;\omega_\textsc{bd}<-69.95,$ or $\omega_\textsc{bd}>79.44\;|\;\omega_\textsc{bd}<-81.44,$ depending on the observational data assumed, which are consistent with the estimates of \cite{bd-coupling, aquaviva, chiva}. These results are to be contrasted with the ones in Ref. \cite{hrycyna}: $\omega_\textsc{bd}=-3/2$, or in \cite{hrycyna-1}: $\omega_\textsc{bd}\approx-1$, which were based on computations made at perturbed solutions.

\section{acknowledgment}

The authors want to thank J D Barrow, S D Odintsov and A Alho, for pointing to us several indispensable bibliographic references. The work of R G-S was partially supported by SIP20150188, SIP20144622, COFAA-IPN, and EDI-IPN grants. I Q thanks CONACyT of M\'exico for support of this research. The authors are grateful to SNI-CONACyT for continuous support of their research activity.


\end{document}